\documentclass[fleqn, usenatbib]{mnras}
\usepackage{epsfig,graphics}
\usepackage{amsmath}
\usepackage{amssymb}
\usepackage{subfloat}
\usepackage{xcolor}
\usepackage{multicol}
\usepackage{ulem}

\title[Intra-night variability of BL~Lac]{Intra-night optical flux and polarization variability of BL~Lacertae during its 2020 $-$ 2021 high state}

\author[Bachev et al.]{Rumen Bachev$^{1}$\thanks{E-mail: bachevr@astro.bas.bg}, Tushar Tripathi$^{2,3}$, Alok C. Gupta$^{4,2}$, Pankaj Kushwaha$^{5}$, \newauthor
Anton Strigachev$^{1}$,  	
Alexander Kurtenkov$^{1}$, Yanko Nikolov$^{1}$, Svetlana Boeva$^{1}$,
\newauthor Goran Damljanovic$^{6}$,  
Oliver Vince$^{6}$, 
Milan Stojanovic$^{6}$,
Shubham Kishore$^{2,3}$, \newauthor
Haritma Gaur$^{2}$, Vinit Dhiman$^{2,7}$, Junhui Fan$^{8,9}$, 
  Nibedita Kalita$^{10}$, 	
Borislav Spassov$^{1}$,  \newauthor
Evgeni Semkov$^{1}$
\\
\\
$^{1}$ Institute of Astronomy and NAO, Bulgarian Academy of Sciences, 1784 Sofia, Bulgaria\\
$^{2}$ Aryabhatta Research Institute of Observational Sciences (ARIES), Manora Peak, Nainital 263001, India\\
$^{3}$ Department of Physics, DDU Gorakhpur University, Gorakhpur 273009, India \\
$^{4}$ Key Laboratory for Research in Galaxies and Cosmology, Shanghai Astronomical Observatory, Chinese Academy of Sciences, Shanghai \\
~~~200030, China \\
$^{5}$Department of Physical Sciences, Indian Institute of Science Education and Research (IISER) Mohali, Knowledge City, Sector 81, \\
~~~SAS Nagar, Punjab 140306, India \\
$^{6}$ Astronomical Observatory, Volgina 7, 11060 Belgrade, Serbia \\
$^{7}$ School of Studies in Physics and Astrophysics, Pt. Ravishankar Shukla University, Amanaka G.E. Road, Raipur 492010, India \\
$^{8}$ Center for Astrophysics, Guangzhou University, Guangzhou 510006, China\\
$^{9}$ Astronomy Science and Technology Research Laboratory of Department of Education of Guangdong Province, Guangzhou 510006, China\\
$^{10}$ Key Laboratory for Polar Science, MNR, Polar Research Institute of China, 451 Jinqiao Road, Shanghai 200135, China}

\begin{document}

\date{Accepted \dots Received \dots; in original form \dots}

\maketitle

\label{firstpage}

\begin{abstract}
	In this work, we report the presence of rapid intra-night optical variations in both -- flux and polarization of the blazar BL Lacertae during its unprecedented 2020--2021 high state of brightness. The object showed significant flux variability and some color changes, but no firmly detectable time delays between the optical bands. The linear polarization was also highly variable in both -- polarization degree and angle (EVPA). The object was observed from several observatories throughout the world, covering in a total of almost 300 hours during 66 nights. Based on our results, we suggest, that the changing Doppler factor of an ensemble of independent emitting regions, travelling along a curved jet that at some point happens to be closely aligned with the line of sight can successfully reproduce our observations during this outburst. This is one of the most extensive variability studies of the optical polarization of a blazar on intra-night timescales. 
    %To the best of our knowledge, this is the first time a blazar has been studied in terms of polarization variability on intra-night time scales to such an extent.
\end{abstract}


\section{Introduction}
BL Lacertae is the archetype of the blazar-class objects, a class of active galactic nuclei, which can generally be characterized by two main features: i) almost entirely non-thermal %\sout{electromagnetic} 
continuum spectrum in the complete electromagnetic (EM) bands, stretching up to 15 orders of magnitude on the frequency scale, with a characteristic broad bimodal spectral energy distribution (SED) \citep[e.g.][]{Foss1998, Abdo2010} and ii) intermittent flux, spectral and polarization variability in all bands and on all time scales, from minutes to years \citep[e.g.,][and references therein]{1989Natur.337..627M,Rait2009,2010A&A...524A..43R,2013MNRAS.436.1530R,2014ApJ...781L...4G,2015A&A...582A.103G,2015MNRAS.452.4263G,2015MNRAS.450..541A,2022MNRAS.509...52D,2022MNRAS.513.4645S,Imaz2022,2022PASJ...74.1041H}. 
The lower energy part of the SED is dominated by synchrotron radiation but the high energy part of SED remains in question. High energy portion of SED is generally explained by leptonic models e.g. SSC (synchrotron self Compton), or EC (external Compton); and alternate hadronic models are preferable for some sources \citep[e.g.,][and references therein]{2019NatAs...3...88G,2013ApJ...768...54B}. From the composite UV/optical spectral perspective, blazars are divided into two classes: BL Lacertae objects (BLLs) and flat spectrum radio quasars (FSRQs), depending on whether they show prominent emission lines and/or thermal accretion disk signatures (latter) or not (former). In addition, an apparent correlation between the frequency at which the two hump of the SED peaks, as well as its correlation with luminosity \citep{Foss1998}, an SED based classification has been argued that designate a blazar based on the frequency ($\nu_{syn}^{peak}$) at which synchrotron part (low-energy hump) peaks.  If $\nu_{syn}^{peak} \leq 10^{14}$ Hz, the blazars are called low synchrotron peaked (LSPs), if $10^{14} \leq \nu_{syn}^{peak} \leq 10^{15}$ Hz then called intermediate synchrotron peaked (ISPs), and if $\nu_{syn}^{peak} \geq 10^{15}$ Hz then known as high synchrotron peaked (HSPs) blazars \citep{Abdo2010}. 
%\sout{their frequency peaks} (the synchrotron \sout{and the inverse-Compton ones}) into High Spectrum (HSP), \red{intermediate (ISP)}  or Low Spectrum (LSP) objects \citep{Abdo2010,Foss1998} \sout{(Abdo et al., 2010)} \red{\bf THIS NEEDS MODIFICATION. STILL NOT CORRECT AND WITHOUT MENTIONING HUMP AND SYNCHROTRON COMPONENT, SYNCHROTRON is introduced}. 
The commonly accepted understanding, however, is that all blazars possess a strong relativistic jet that happens to be pointed almost directly towards the observer’s line of sight \citep{Urry1995}, and which produces most of the energy via non-thermal processes. This energy, covering practically the entire observable electromagnetic spectrum is additionally Doppler-boosted towards the observer, as being generated along a relativistic jet, which significantly amplifies the fluxes and shortens the characteristic variability timescales. Therefore it is not surprising to observe significant (say 10\%) optical/gamma variations for as short as $\sim$ 15 min occasionally in some of these objects \citep[e.g.][]{Ulr1997}. \\ %\sout{(e.g. Ulrich, Maraschi \& Urry, 1997)}. 
\\
It is not yet clear why, but the FSRQs (and perhaps some LSP BLLs) statistically turn out to be more variable on both -- intra-night and long-term time scales in optical bands, than the HSP BLLs, as the former show faster and more prominent variations \citep[e.g.][and references therein]{Bach2018}. The explanation of this tendency, if further confirmed, could probably be looked for in terms of the jet producing mechanisms -- being based on the extraction of energy of the black hole spin \citep{Blan1977} or on the accretion disk itself \citep{Blan1982}.\\
\\
BL Lacertae, located at a cosmological redshift of z=0.069 \citep{Oak1974, Macl1968, Schm1968}, %was identified as a star-like 
is a prototype of BL~Lacertae class of sources,  that were
first identified as sources with peculiar spectra and strong and rapid flux variability in radio and optical bands \citep[e.g.,][and references therein]{1989Natur.337..627M,Fan1998,Moor1982,Thor2002,2015MNRAS.450..541A}. The perplexing behavior of flux variations on timescales of a few hours to within a day, now widely referred in the literature as intra-night (or intra-day, IDV) variability \citep{1995ARA&A..33..163W} feature has been reported even in the very early optical observations \citep[e.g.][]{Raci1970,DuPu1969,1989Natur.337..627M}, and has now become one of the identifying criteria for blazars. Since then, the source has been extensively observed and explored source across the EM spectrum, independently as well as in a coordinated ways \citep{1989Natur.337..627M,Fan1998,Rait2009,2010A&A...524A..43R,2013MNRAS.436.1530R,2015A&A...582A.103G,2015MNRAS.452.4263G,2015MNRAS.450..541A,MAGI2019,Weav2020}.
%in order to understand and establish characteristic features of such sources. with peculiar spec. the criteria of The  the primary driver of more observations in different energy bands. 
These studies reveals not only strong flux and polarization variability, but significant spectral changes as well, especially in the X-ray band \citep[e.g.][]{Rait2009}. The anti-correlated optical flux and degree of polarization is detected in the source in 2008 -- 2009 observing season \citep{2014ApJ...781L...4G}. In the historic and unprecedented optical activity of 2020--2021 \citep{Marc2021,Kunk2021}, an extremely
soft X-ray spectrum was seen \citep[e.g.][and referecnes therein]{Prin2021}, a transition similar to that of
another BL~Lac object OJ~287 \citep[e.g.][]{Kush2022,Kush2018}. This historic activity was followed in all the EM
bands \citep[e.g.][and references therein]{Marc2021, Jors2022, Prin2021}, and we too monitored the source optical brightness and polarization intensively and extensively. Recently a $\sim$ 13 hours optical quasi-periodic oscillation is reported in BL Lacerate in its observations in the historical pre-outburst in 2020 \citep{Jors2022}. In another recent optical flux and spectral variability study carried out during this historical outburst, IDVs with amplitude up to $\sim$ 30\% were detected, and the spectral evolution predominated
by flattening of the spectra with increasing brightness i.e., a bluer-when-brighter trend was observed \citep{2022arXiv221204181K}. \\
\\
In this paper we present the intra-night optical (flux and polarimetric) variability study of BL Lacertae during its year 2020--2021 optical maximum, unsurpassed for at least 15 years (see e.g. Tuorla monitoring program\footnote{\url{https://users.utu.fi/kani/1m/}}, St. Petersburg monitoring program\footnote{\url{https://vo.astro.spbu.ru/en/program}}, etc.). The object reached $R\simeq11.5$ mag at some point \citep{Marc2021,Kunk2021}, a historical maximum which has never been reported before, at least what concerns the last decades. The studies in the optical region are important, as often there the synchrotron peak is located (at least for LSP's). The strong intra-night variability is a well-documented and studied feature in many blazars, including BL Lacertae \citep[e.g.,][and references therein]{1989Natur.337..627M,Papa2003,Stal2006,Zhai2012,2015MNRAS.452.4263G,2015MNRAS.450..541A,Bhat2018,Fang2022,Imaz2022}. \\
\\
Despite the significant progress throughout the years, yet no entirely consistent picture has emerged to be able to account for all observational signatures. On the other hand, the physics of the relativistic jets can be really complicated and can include processes like standing or propagating shocks; blobs or ``plasmoids", moving along the jet; relatively ordered (helical) or highly disordered (turbulent) magnetic fields, etc., all of which can, one way or another, influence the synchrotron emission, even on intra-night time scales. In addition, the Doppler boosting factor is another major ``modifier" of the emission, capable of significantly (and non-linearly) increasing the flux intensity in the observer's frame of reference  \citep[e.g.,][and references therein]{2013ApJ...768...54B,2014ApJ...780...87M,2015IAUS..313..122M,2014ApJ...789...66Z,2015ApJ...804...58Z}. Therefore, a systematic study of the intra-night variability of a blazar can significantly help to improve the existing models, especially when color and polarimetric variations are also simultaneously investigated. 
Our main goal will be to study how during the maximum light of BL lacerate the intra-night activity changes with the average brightness in the optical band. We search for intra-night color changes, study changes in the polarization rate and the orientation of the electric vector (EVPA), search for relations among different temporal characteristics of the intra-night variations, etc. 
To the best of our knowledge, this is one of the first time the polarimetric variability of this blazar is systematically studied on intra-night time scales.  
Recently, similar studies exploring intra-night photo-polarimetric variability of BL~Lacerate in which the source has also shown significant flux and polarimetric variations were reported \citep{Imaz2022, Shablo2022}. \\
\\
In this paper we concentrate only on the short-term (intra-night) variability of BL~Lacertae. The long-term flux and polarization variability results, covering this exceptional period will be published elsewhere (Raiteri et al., 2023, to be published). \\
\\
The paper is structured as follows. In section 2, we provide observations carried out from different telescopes and their data analysis. Results are reported in Section 3. The Discussion and Conclusion of the manuscript is provided in Section 4 and Section 5, respectively. \\
%\\
%\blue{\bf PANKAJ and HARITMA, please carefully read the Introduction Section and if any modification is needed, do that.}

%Similar study on intra-night variability of BL Lacertae during the last months of 2019 was reported by \citet{Fang2022} \sout{Fang et al. (2022)} \red{and \citep{Imaz2022}}. \red{In the former the} \sout{These} authors find rapid intra-night variations, color changes and occasional time delays between the optical bands, results similar to those we report in this paper.

\section{Observations and Data Analysis}
BL Lacertae was monitored for intra-night variability during 66 nights for a total of almost 300 hours throughout the years 2020--2021. Thus, the average monitoring duration was about 4 hours, but ranged from about 1 to more than 8 hours. In addition, data from other 24 nights (about 73 hours), 
taken between years 2017 and 2019, when the blazar was not in a maximum state, were also analyzed. %Table \ref{tab:log} presents {\bf a sample of} the log of the observations. {\bf The complete observations log is presented in Table \ref{appendix:A1}.} 
A sample of log of observations is shown as Table \ref{tab:log}, whereas a complete list is presented in the appendix as Table \ref{appendix:A1}.\\

\begin{table*}
	
	\caption{A sample of Observational log. The complete observation log is given in Table \ref{appendix:A1}} 
	\label{tab:log} 
	\centering 
	%\resizebox{\textwidth} {!}{ 
	\begin{tabular}{lrcccccc} 	\hline\hline 
		JD	&	Evening	&	Telescope	&	Duration	&	Filters	&	$<R>$	&	$\sigma( R )$	&	$<V-I>$	\\\hline 															
		2458011.35	&	14.9.17	&	B60	&	4.6	&	BVRI	&	13.150	&	0.020	&	1.45	 \\
		2458012.35	&	15.9.17	&	B60	&	4.0	&	BVRI	&	13.250	&	$<$0.005	&	1.44	 \\
		2458341.46	&	10.8.18	&	B60	&	2.7	&	BVRI	&	12.900	&	0.010	&	1.38	 \\
		2458343.44	&	12.8.18	&	B60	&	2.5	&	BVRI	&	13.500	&	0.010	&	1.41	 \\
		2458344.43	&	13.8.18	&	B60	&	4.0	&	BVRI	&	12.940	&	0.020	&	1.38	 \\\hline
		\end{tabular}
\end{table*}

\noindent
The observations were performed with several small and middle-class telescopes, such as 2-m RCC telescope of Rozhen National Observatory\footnote{\url{https://www.nao-rozhen.org/}} (Bulgaria), 0.6-m telescope of Belogradchik Observatory\footnote{\url{https://astro.bas.bg/AOBel/index.php}} (Bulgaria), 1.3-m telescope of \footnote{\url{https://skinakas.physics.uoc.gr/en/}} (Greece), 1.04-m telescope of ARIES\footnote{\url{https://www.aries.res.in/facilities/astronomical-telescopes/104m-telescope}}, Nainital (India), 1.4-m and 0.6-m telescope of Vidojevica Observatory\footnote{\url{http://vidojevica.aob.rs/}} (Serbia). All telescopes were using CCD’s, equipped with standard (Johnson-Cousins) UBVRI broad band filter sets. Additionally, the Belogradchik 0.6-m telescope was equipped with a double-barrel filter set, made by FLI, allowing a combination of UBVRI with polarimetric filters, thus making possible to perform a linear polarimetry of the blazar in the optical band selected. We also used the FoReRo-2 mode of the 2-m Rozhen telescope, allowing a beam-splitting between the red and the blue channels, thus allowing simultaneous exposures in both channels, using two separate (blue- and red-sensitive) CCD’s.

All frames  thus collected were properly reduced (flat field, bias, dark frame -- where applicable) and an aperture photometry was performed in order to extract the magnitudes. The photometric measurements were made with respect to the standard stars B, C and H with an aperture radius of 8 arcsec, as suggested by the GASP consortium\footnote{\url{https://www.oato.inaf.it/blazars/webt/2200420-bl-lac/}}, and the BVRI magnitudes of these standard stars are taken from \citet{1969A&A.....3..436B}  and \citet{1996A&AS..116..403F}.
The (quasi)-simultaneous multicolor observations were made via taking BVRI frames in a consecutive order (Skinakas, Belogradchik and Vidojevica) or by using a beam splitter, allowing simultaneous BR observations (Rozhen). Note that for an exposure time of 60 -- 120 sec, the combined time resolution of the light curves in the former approach could not be shorter than 5 minutes and was typically less than a minute in the latter. The details about telescopes, CCDs, filter systems, and photometric data analysis are provided in our series of papers \citep[e.g.,][]{2012MNRAS.420.3147G, Strig2011,2012MNRAS.424.2625B,Bach2017,2017MNRAS.465.4423G,2019AJ....157...95G,2020ApJ...890...72P}. \\

\begin{table*}
	
	\caption{Polarimetric results (observations only in $R$-band.)} 
	\label{tab:polarimetry} 	
    \resizebox{7in}{!}{
	\begin{tabular}{crccccccccc} 
		\hline\hline 

JD	&	Evening	&	Duration	&	$<p>$	&	$<Err(p)>^\ast$	&	$\sigma(p)$	&	Probability	&	$<EVPA>$	&	$<Err(EVPA)>$	&	$\sigma(EVPA)$	&	Probability	\\      &  dd.mm.yy & (hours)       & \%        & \%            &               &   \%          & degree   & 
  degree    &                   & \% \\  \hline

2459112.27	&	19.9.20	&	$-$	&	9.3	&	0.5	&	$-$	&	$-$	&	30.4	&	1.7	&	$-$	&	$-$	 \\
2459113.29	&	20.9.20	&	$-$	&	7.9	&	0.7	&	$-$	&	$-$	&	24.5	&	2.2	&	$-$	&	$-$	 \\
2459114.28	&	21.9.20	&	$-$	&	11.2	&	0.3	&	$-$	&	$-$	&	23.6	&	0.6	&	$-$	&	$-$	 \\
2459115.30	&	22.9.20	&	$-$	&	8.4	&	1.1	&	$-$	&	$-$	&	26	&	0.8	&	$-$	&	$-$	 \\
2459116.35	&	23.9.20	&	$-$	&	8	&	0.4	&	$-$	&	$-$	&	27.9	&	1.4	&	$-$	&	$-$	 \\
2459117.28	&	24.9.20	&	$-$	&	6.7	&	0.3	&	$-$	&	$-$	&	12	&	1.8	&	$-$	&	$-$	 \\
2459134.25	&	11.10.20	&	$-$	&	4.8	&	0.4	&	$-$	&	$-$	&	24.8	&	4.3	&	$-$	&	$-$	 \\
2459141.29	&	18.10.20	&	$-$	&	8.2	&	0.8	&	$-$	&	$-$	&	6.7	&	2.9	&	$-$	&	$-$	 \\
2459142.29	&	19.10.20	&	$-$	&	4.5	&	0.5	&	$-$	&	$-$	&	1.8	&	3.8	&	$-$	&	$-$	 \\
2459143.29	&	20.10.20	&	$-$	&	6.7	&	0.7	&	$-$	&	$-$	&	167.9	&	2.1	&	$-$	&	$-$	 \\
2459144.24	&	21.10.20	&	$-$	&	6.1	&	0.8	&	$-$	&	$-$	&	29.2	&	4.1	&	$-$	&	$-$	 \\
2459145.26	&	22.10.20	&	$-$	&	10.1	&	0.5	&	$-$	&	$-$	&	10.2	&	1.3	&	$-$	&	$-$	 \\
2459161.44	&	7.11.20	&	$-$	&	11.6	&	0.8	&	$-$	&	$-$	&	12.9	&	2.2	&	$-$	&	$-$	 \\
2459386.48	&	20.6.21	&	$-$	&	5.9	&	0.4	&	$-$	&	$-$	&	57.8	&	2.2	&	$-$	&	$-$	 \\
2459403.51	&	7.7.21	&	2.4	&	15.2	&	0.7	&	2.0	&	$>$99.999	&	11.2	&	1.5	&	1.4	&	$>$90	 \\
2459404.51	&	8.7.21	&	2.3	&	12.9	&	0.4	&	0.5	&	$<$90	&	10.9	&	1.1	&	0.4	&	$<$90	 \\
2459406.52	&	10.7.21	&	1.9	&	5.1	&	0.5	&	1.0	&	$>$99.9	&	9.1	&	2.1	&	3.3	&	$>$99	 \\
2459407.49	&	11.7.21	&	4.0	&	8.4	&	0.6	&	1.5	&	$>$99.999	&	9.6	&	2.6	&	5.6	&	$>$99.999	 \\
2459408.42	&	12.7.21	&	0.5	&	16.2	&	1.1	&	2.3	&	$>$95	&	0.8	&	1.0	&	0.6	&	$<$90	 \\
2459434.45	&	7.8.21	&	5.6	&	5.2	&	0.4	&	1.1	&	$>$99.999	&	178.2	&	3.0	&	2.7	&	$>$90	 \\
2459435.46	&	8.8.21	&	5.7	&	12.7	&	0.5	&	1.1	&	$>$99.999	&	10.7	&	1.3	&	2.2	&	$>$99.999	 \\
2459436.50	&	9.8.21	&	5.5	&	8.2	&	0.9	&	1.5	&	$>$99.99	&	18.9	&	3.3	&	6.8	&	$>$99.999	 \\
2459437.45	&	10.8.21	&	5.9	&	2.5	&	0.4	&	0.9	&	$>$99.999	&	2.5	&	6.8	&	7.0	&	$>$90	 \\
2459438.43	&	11.8.21	&	5.1	&	8.9	&	0.6	&	1.8	&	$>$99.999	&	156.6	&	1.8	&	3.7	&	$>$99.999	 \\
2459439.44	&	12.8.21	&	5.4	&	7.2	&	0.4	&	0.9	&	$>$99.999	&	146.9	&	1.8	&	1.9	&	$>$99	 \\
2459440.42	&	13.8.21	&	5.6	&	7.2	&	0.5	&	0.6	&	$>$99.9	&	149.3	&	1.9	&	2.9	&	$>$99.999	 \\
2459443.54	&	16.8.21	&	2.0	&	4.4	&	0.5	&	1.0	&	$>$99.99	&	160	&	4	&	9.0	&	$>$99.999	 \\
2459468.42	&	10.9.21	&	8.0	&	4.2	&	0.4	&	0.7	&	$>$99.999	&	87.2	&	3.4	&	9.2	&	$>$99.999	 \\
2459469.43	&	11.9.21	&	7.8	&	8.4	&	0.4	&	1.2	&	$>$99.999	&	56.9	&	1.9	&	2.8	&	$>$95	 \\
2459470.43	&	12.9.21	&	8.1	&	10.4	&	0.5	&	1.0	&	$>$99.999	&	69.7	&	1.5	&	3.9	&	$>$99.999	 \\
2459471.42	&	13.9.21	&	8.1	&	4.3	&	0.6	&	0.9	&	$>$99.999	&	72.6	&	4.4	&	16.3	&	$>$99.999	 \\
2459472.40	&	14.9.21	&	6.3	&	6.2	&	0.8	&	0.6	&	$>$99	&	85.4	&	3.6	&	5.7	&	$>$99.999	 \\
2459473.40	&	15.9.21	&	6.9	&	4.6	&	0.5	&	1.1	&	$>$99.999	&	108.4	&	3.0	&	9.7	&	$>$99.999	 \\
2459520.34	&	1.11.21	&	6.4	&	14.5	&	0.4	&	0.5	&	$>$95	&	30.3	&	0.6	&	0.8	&	$>$99.9	 \\
2459604.21	&	24.2.22	&	$-$	&	11.6	&	0.5	&	$-$	&	$-$	&	26.3	&	1.1	&	$-$	&	$-$	 \\
\hline 
\end{tabular}}
$^\ast$ Standard deviations, as mentioned in Sect. 2.
\end{table*}

\noindent
Three polarimetric filters (oriented at 0--180, 60--240 and 120--300 degrees) were used to obtain the polarization degree ($p$) and the Electric Vector Polarization Angle (EVPA). This approach obviously could not employ the standard Stokes parameters and requires instead solving 3 equations for 3 unknowns \citep{Bach2023}:
\\
\\
%\begin{array}{l}
$I_{\rm 0}=\frac{1}{2}I_{\rm np}+I_{\rm p}\cos^{2}\theta$ \\
$I_{\rm 60}=\frac{1}{2}I_{\rm np}+I_{\rm p}\cos^{2} (\theta-60)$ \\
$I_{\rm 120}=\frac{1}{2}I_{\rm np}+I_{\rm p}\cos^{2} (\theta-120)$, where\\
$p[\%]=100\frac{ I_{\rm p}}{ I_{\rm np}+ I_{\rm p}}$, and $EVPA = \theta$
%\end{array}
\\
\\
We solved these equations numerically. For each orientation at least 3, but normally 5 frames were taken, in a consecutive order, and the standard deviation of the object's magnitude among these frames was taken as a proxy of the photometric error. Note that any intrinsic variability (the typical polarization exposure time was 120 sec) in between the frames will only artificially increase this error. Unfortunately, this is the limitation of our apparatus. The errors of $p$ and EVPA were obtained by varying the obtained magnitudes in each orientation within their respective photometric errors. For the magnitudes a Gaussian distribution was assumed, which is approximately correct; however, the polarization parameters are certainly not Gaussian distributed ($p$ is not even symmetrically distributed), i.e. the errors presented in Fig. \ref{fig:f6} and Table  \ref{tab:polarimetry} are just the standard deviations. All polarimetric measurements were made in \textit{R}-band only. %\blue{\bf RUMEN, please add some references which describe the used polarimetric systems. I don't have information about it.}- done above

\section{Results}

\subsection{Intra-night variability}
\begin{figure*}
\centering
	\includegraphics[width=100mm]{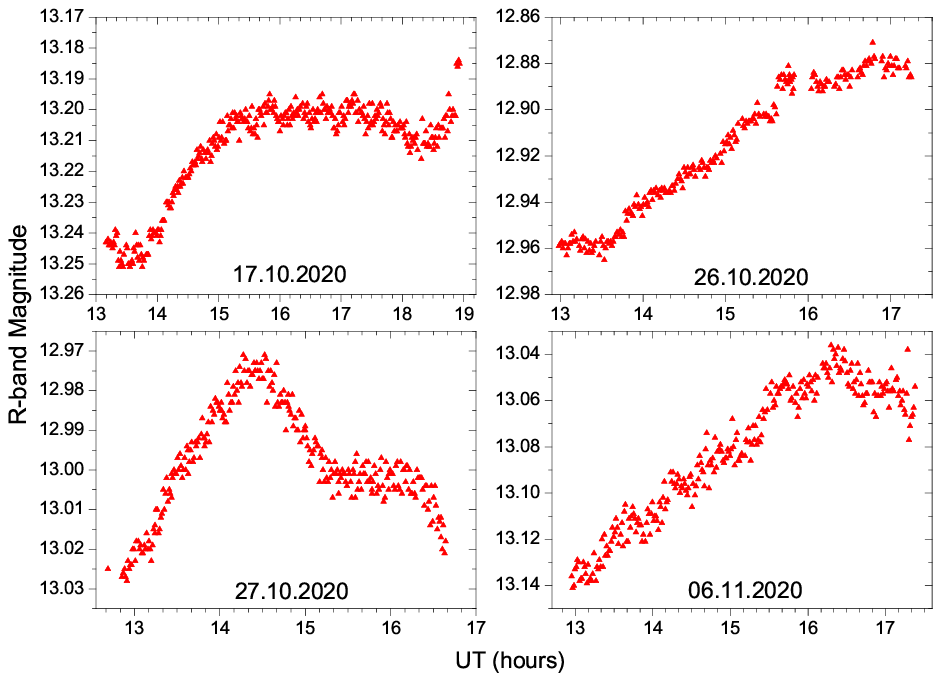}
	\caption{Examples of \textit{R}-band intra-night monitoring with the 1m ARIES telescope.}
	\label{f4}
\end{figure*}

Tables \ref{tab:log}, \ref{appendix:A1} and \ref{tab:polarimetry} summarize the results of our optical photometric and polarimetric monitoring of BL Lacertae.  The figures (Fig. \ref{f1} -- \ref{f4}) show examples of prominent variations on intra-night timescales. The upper panel of each figure shows the light curves in one or more filter bands for the different instruments. The magnitudes of some bands have been arbitrarily shifted for presentation purposes. Again, for presentation purposes no photometric errors were shown; however, during this very bright state, they all were typically $<0.01$ mag ($<0.02$ mag for B-band, for the smallest Belogradchik and Vidojevica telescopes). In other photometric bands i.e. VRI, we have even lesser errors than B-band. The evening date of each observation is indicated. \\
\\
One sees significant intra-night variability, at least during the periods of high flux states. All bands show unpredictable, but systematic behavior among them. Variations sometimes reach up to 0.3 mag within several hours of observations (e.g. the night of 21.09.2020, Fig. \ref{f1a}), and can be as rapid as of $\sim$0.1 mag for half an hour (e.g. 28.08.2021, Fig. \ref{f2a}). To explore
intra night variability, we exploited two of the most widely used methods: Power enhanced F-test \citep{2014AJ....148...93D, 2015AJ....150...44D} and Nested-ANOVA \citep{1998ApJ...501...69D,2015AJ....150...44D} as detailed below. An intra-night LC is reported to be variable only if it is found variable by both, enhanced F-test and nested ANOVA test.

\subsubsection{Power enhanced F-test}
The Power enhanced F-test is introduced by \citet{2014AJ....148...93D} 
 and \citet{2015AJ....150...44D}. We use one comparison star as reference star to find the differential light curves (DLCs) of the blazar and other comparison stars. The variance of blazar light curve (LC) is compared with the combined variance of comparison stars. It is defined as \citep{Pand2019} 
\begin{equation}
F_{enh}=\frac{s_{blz}^2}{s_c^2}
\end{equation}

\noindent
where, $s^2_{blz}$ is the variance of the DLCs of difference of instrumental magnitude of blazar and reference star, and
\begin{equation}
s_c^2=\frac{1}{\left(\sum_{j=1}^k N_j\right)-k} \sum_{j=1}^k \sum_{i=1}^{N_i} s_{j, i}^2
\end{equation}

\noindent
and is defined as the variance of the combined Differential LCs of difference of instrumental magnitude of comparison star and reference star. $N_{j}$ is the number of data points of the $j$-th comparison star, and $s^2_{j,i}$ is its scaled square deviation, which is defined as           
\begin{equation}
s_{j, i}^2=\omega_j\left(m_{j, i}-\bar{m}_j\right)^2
\end{equation}

\noindent
where $\omega_j, m_{j, i}$ and $\bar{m}_j$ are scaling factor of $j$-th comparison star DLC, its differential magnitude and its mean magnitude respectively. The averaged square error of the blazar Differential LC divided by the averaged square error of the $j$-th comparison star is used as the scaling factor. \\
\\
In this work, we have 3 comparison stars B, C, and H. As BL~Lacertae magnitude was closest to the comparison star C so we have taken C as the reference star. Since, two more comparison stars are left so $k=2$. The blazar and all the comparison stars have the same number of observations $N$, so the degree of freedom in the numerator is $(N-1)$ and the denominator is $k(N-1)$. We have found the $F_{enh}$ using equation (1) and compared it with $F_{c}$ at the confidence level of 99\% i.e. $\alpha$ = 0.01. If $F_{enh} > F_{c}$  then DLC is considered as Variable (V) otherwise it is considered as Non-Variable (NV).

\subsubsection{Nested-ANOVA Test}
For AGN variability, the one-way analysis of variance (ANOVA) test was introduced by \citet{1998ApJ...501...69D}. The nested ANOVA test is an updated version of the ANOVA test \citep{2015AJ....150...44D}.  In nested ANOVA test, we use all the comparison stars as reference stars to find the differential LCs. Unlike power enhanced F-test which needed one comparison star, here we are using all the comparison stars as reference stars, so we have one more star to work with. We have three comparison stars in this work namely B, C, and H to generate the differential LCs of the blazar. We group these differential LCs in such a way that we have 5 points in each group. From equation (4) of \citet{2015AJ....150...44D}, we have estimated the value of $MS_G$ (mean square due to groups) and $MS_{O(G)}$ (mean square due to nested observations in groups). We then estimated the F-statistics using the ratio $F=MS_G/MS_{O(G)}$. At a confidence level of 99\% i.e $\alpha$=0.01, if F-statistics $> F_{C}$ then we say that LCs are variable (V) otherwise we say it is non-variable (NV). The  sample and detailed results of Power enhanced F-test and Nested-ANOVA test are given in Table \ref{tab:resIDV}  and Table \ref{appendix:B1}, respectively.
%\blue{\bf TUSHAR and VINIT, check section 3.1.1 and 3.1.2 carefully. I feel some papers should be cited in these 2 subsections.}

\begin{table*}
	    \centering
	    \caption{{\bf A sample result of IDV of BL Lac. The complete results of IDV is provided in Table \ref{appendix:B1}}}
	    \begin{tabular}{ccccccccc} 
	   \hline\hline
Obs. date & Obs. start time    & Band &  \multicolumn{2}{c}{ Power enhanced F-test}& \multicolumn{2}{c}{Nested ANOVA} & Variability & $A$\\
          
dd-mm-yyyy &       JD     & & DoF($\nu_1,\nu_2$) & $F_{enh}$ / $F_c$ & DoF($\nu_1,\nu_2$) & F /  $F_c$   & Status & (\%) \\
	    \hline
2017-09-14 & 2458011.26980 & V & 38,76 & 2.27/1.87 &  9,30 & 38.47/3.06 & V & 8.22  \\ 
           & 2458011.26539 & R & 39,78 & 4.12/1.86 &  9,30 & 41.78/3.06 & V & 7.84   \\ 
           & 2458011.26685 & I & 40,80 & 2.26/1.84 &  9,30 & 20.84/3.06 & V & 8.86    \\\hline 
	          \end{tabular}
  \label{tab:resIDV}
\end{table*}          

\subsubsection{Intra-night Variability Amplitude}
For each of the intra-night variable LCs, we estimated the percentage of variability amplitude (A), using the equation provided by \citep{1996A&A...305...42H}.
\begin{equation}
\label{sec:Intra}
A = 100\times \sqrt{(A_{max}-A_{min})^2 - 2 \sigma^2},
\end{equation}
where $A_{max}$ and $A_{min}$ are the maximum and minimum magnitudes in a variable LC, respectively, and $\sigma$ is the mean error of the LC. The amplitude of variability is reported in the last column of Table \ref{tab:resIDV}  and Table \ref{appendix:B1}. 
 
\subsection{A light curves comparison}
\begin{figure}
	\includegraphics[width=90mm]{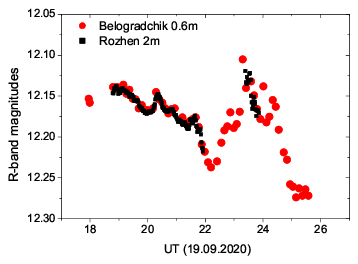}
	\caption{Comparison of the \textit{R}-band LC of BL~Lacertae, obtained with two different instruments, equipped with different CCD’s and filter sets. The close match seen in the figure is indicative for the reliability of our photometry.}
	\label{f5}
\end{figure}

Fig. \ref{f5} presents a comparison of two \textit{R}-band measurements, obtained by two different telescopes (Rozhen 2-m and Belogradchik 0.6-m), during the same night -- 20.09.2020. Clearly the match is impressive, even taking into account the different equipments used in these telescopes, the different altitudes and probably -- atmospheric conditions, etc.; no shifts were needed to match the light curves. Such a comparison allows judging on the reliability of our results.

\subsection{Color changes}
The changes of color (directly related to the source spectrum) are a well-documented feature of the blazars' optical SED. BL lac -- type objects, as BL Lacertae are known to show a general ``bluer-when-brighter, BWB" behavior \citep[e.g.,][and references therein]{2006A&A...450...39G,Gaur2012,Bhat2018}. Color changes on the intra-night time scales are rarely studied \citep[e.g.,][]{2015MNRAS.450..541A} but a genuine color change has been rarely reported for this object \citep[e.g.,][]{2022arXiv221204181K, Imaz2022}. 
Our results clearly demonstrate the presence of a low-level color changes, for a period of a few hours. The second panels of Fig. \ref{fig:f3} show the color variability.  It is clearly seen during some nights (e.g. 21.10.2020, 03.09.2021, etc.). Fig. \ref{f7} indicates a BWB behavior for the averaged values during each observing night, which, as mentioned above is expected for this type of objects.

\begin{subfigures}
\label{f1}
\begin{figure*}
	\includegraphics[width=180mm]{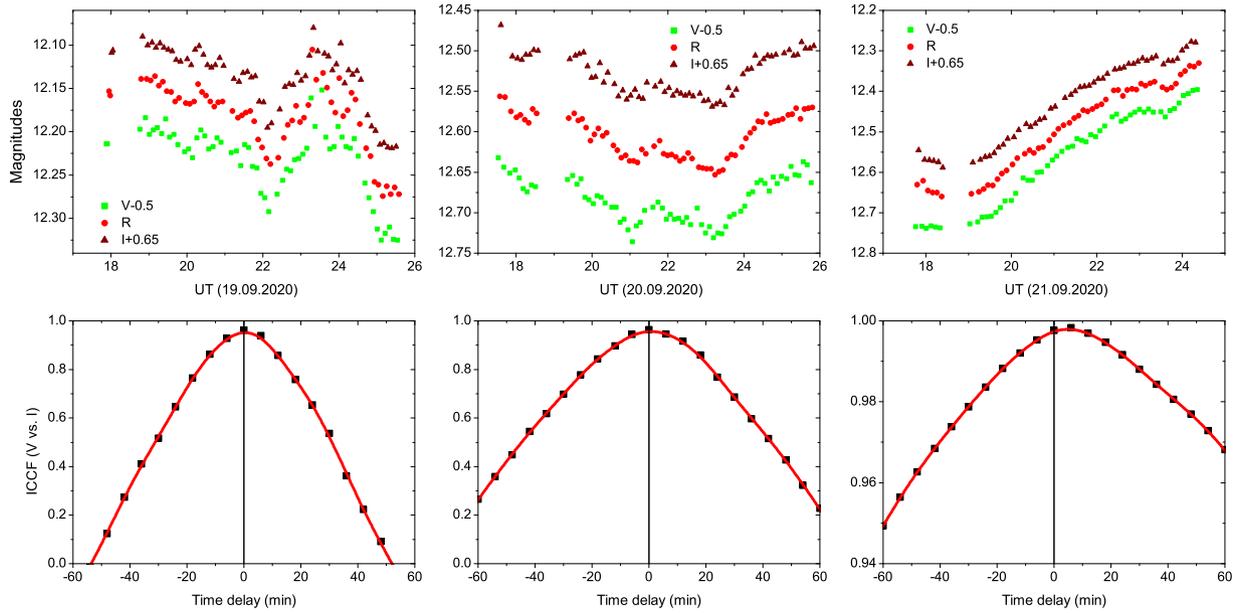}
	\caption{Quasi-simultaneous (repeating \textit{VRI} filters) intra-night light curves of BL Lacertae with the 60 cm Belogradchik telescope. Each lower panel shows the ICCF between \textit{V} and \textit{I} bands (positive time delay means that \textit{V} is leading). All pronounced ICCF peaks appear to coincide with zero lag (at least within the time resolution of about 6 min for this instrument).}
	\label{f1a}
\end{figure*}

%\addtocounter{figure}{-1}
\begin{figure*}
	\includegraphics[width=180mm]{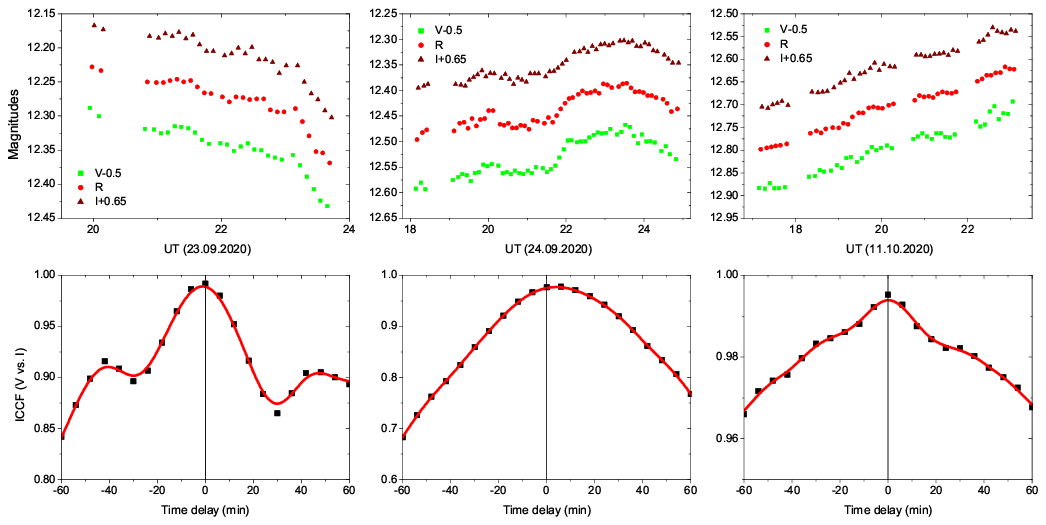}
	\caption{See Figure \ref{f1a}.}
	\label{f1b}
\end{figure*}

%\addtocounter{figure}{-1}
\begin{figure*}
	\includegraphics[width=180mm]{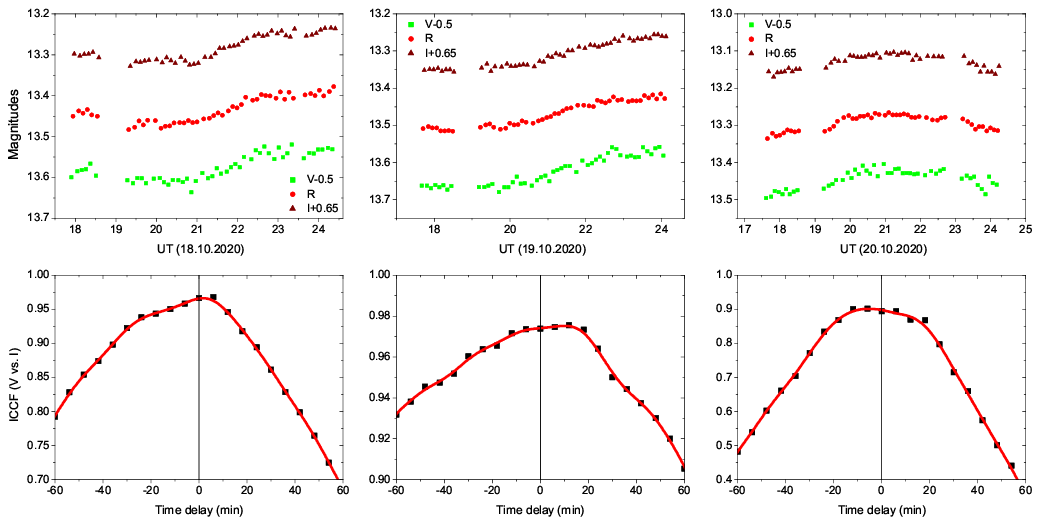}
	\caption{See Figure \ref{f1a}.}
	\label{f1c}
\end{figure*}
\end{subfigures}

\begin{subfigures}
\label{f2}
\begin{figure*}
	\includegraphics[width=180mm]{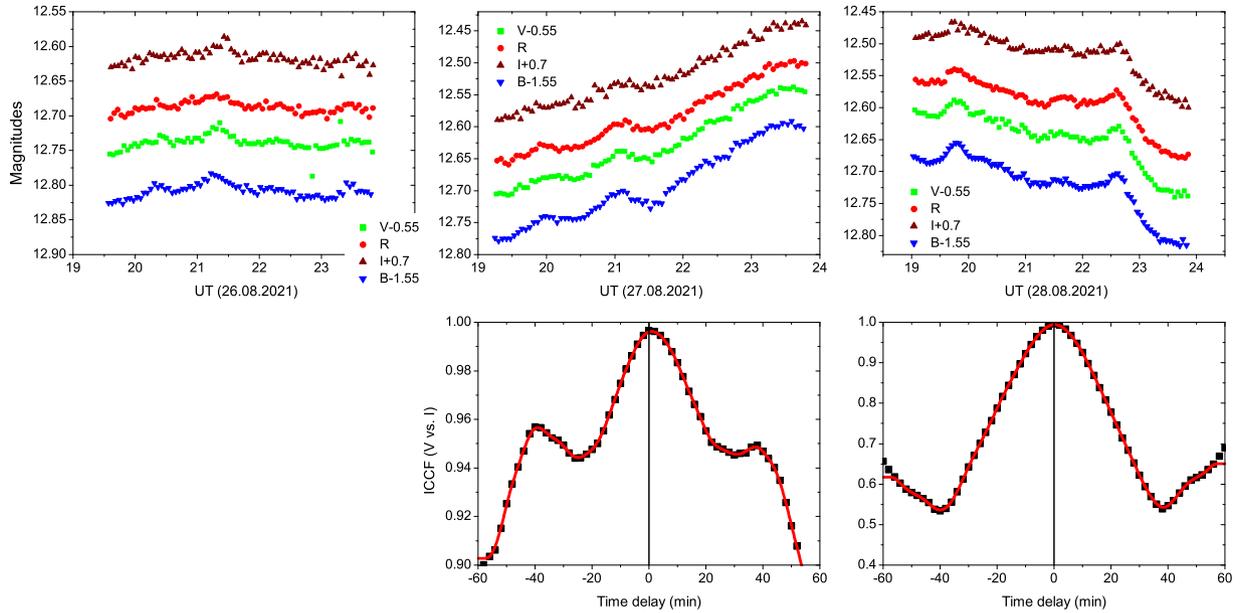}
	\caption{Examples of quasi-simultaneous intra-night monitoring (\textit{BVRI}) with the 1.3m Skinakas telescope. Due to the lack of prominent variability, no cross-correlation of the LCs has been attempted for the night of 26.08.21. Here \textit{B} vs. \textit{R} are used to search for time delays. No such are evident from our data.}
	\label{f2a}
\end{figure*}

%\addtocounter{figure}{-1}
\begin{figure*}
	\includegraphics[width=180mm]{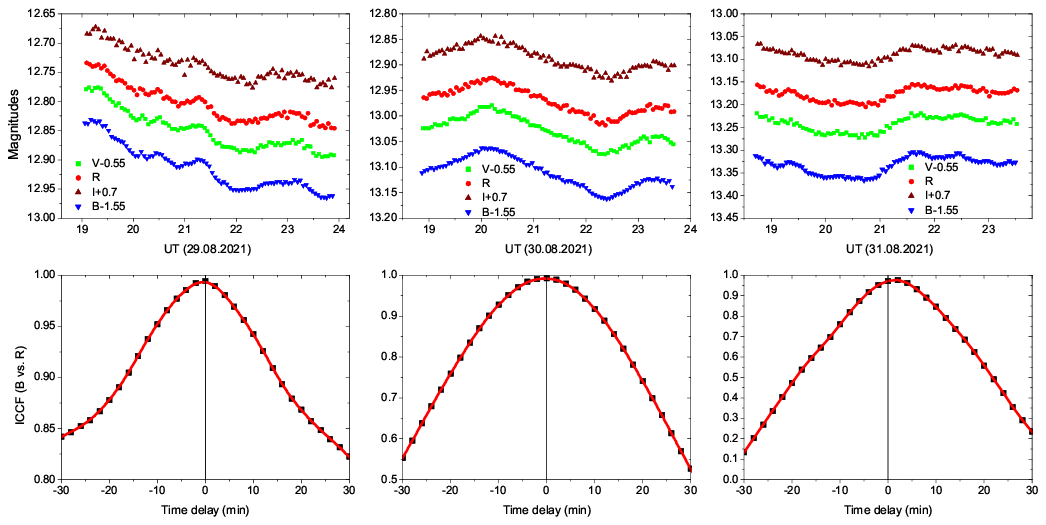}
	\caption{See Figure \ref{f2a}.}
	\label{f2b}
\end{figure*}
\end{subfigures}

\begin{subfigures}
\label{fig:f3}
\begin{figure*}
	\includegraphics[width=180mm]{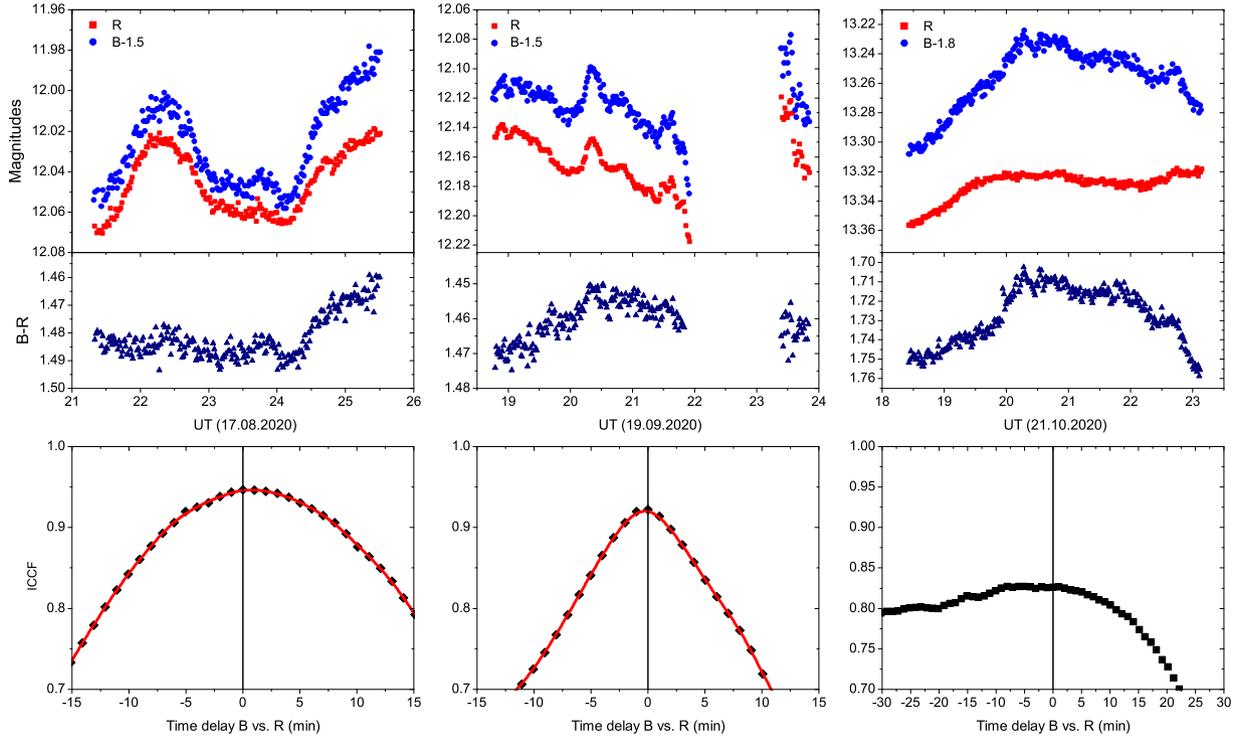}
	\caption{Examples of simultaneous \textit{BR} monitoring with 2m Rozhen telescope. Clearly, BL Lacertae shows short-lasting small-scale micro variations of one or two hundreds of a magnitude, well visible in both bands. The (\textit{B-R}) color (the middle panel of each night) also shows small-scale changes, which do not seem to be highly correlated with the flux level. The ICCFs are generally peaking at zero (within a minute), at least when the peaks are well-defined.}
	\label{f3a}
\end{figure*}

%\addtocounter{figure}{-1}
\begin{figure*}
	\includegraphics[width=180mm]{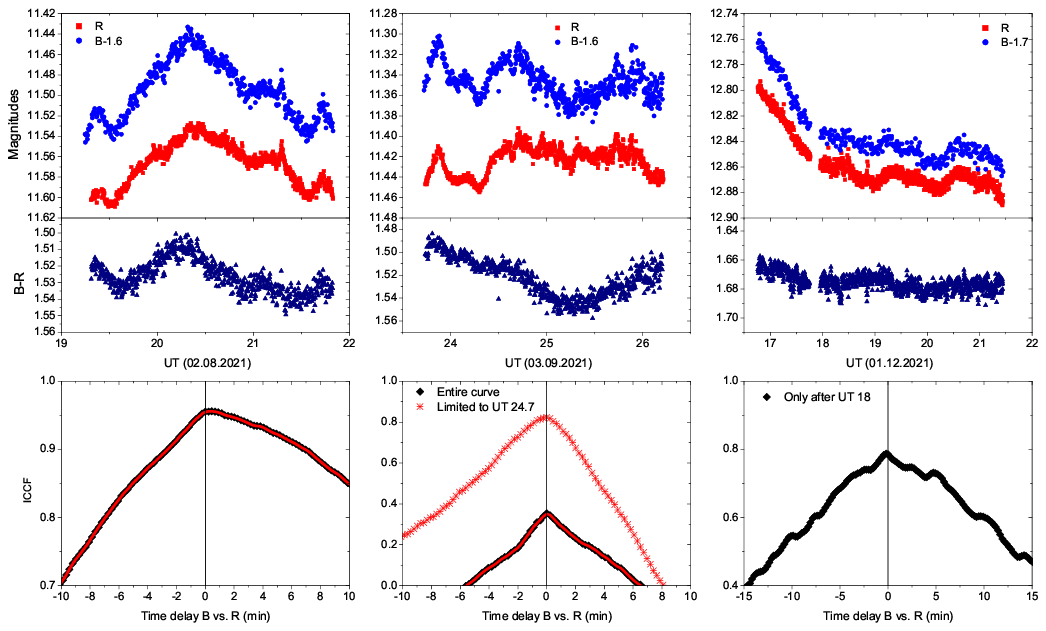}
	\caption{See Figure \ref{f3a}.}
	\label{f3b}
\end{figure*}

%\addtocounter{figure}{-1}
\begin{figure}
	\includegraphics[width=60mm]{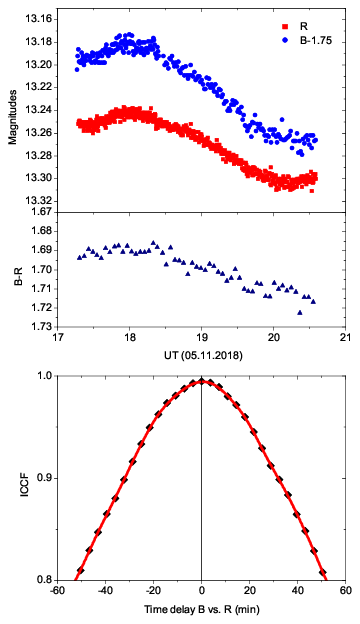}
	\caption{See Figure \ref{f3a}.}
	\label{f3c}
\end{figure}
\end{subfigures}

\subsection{Time lags}
Multi-band monitoring allows searching for possible time delays among the bands. Unfortunately, repeating exposures in consecutive filters limits the time resolution of possible time delay detection. Any detected delays, shorter than the time between two image frames of the same band, should be considered unreliable. In our monitoring, the typical data recording at Belogradchik 0.6-m was $\sim$ 6 min, Skinakas 1.3-m $\sim$ 3 min, Rozhen 2-m $\sim$ 1 min. The other instruments either did not observe in multiple bands or  lacked significant variability to perform this test.
\\
\\
To search for time delays, we employed the Interpolated Cross-Correlation Function (ICCF), a method proposed by \citet{Gask1986} and widely used afterward to study the correlations between unevenly sampled time series data. This method uses a linear interpolation between the adjacent data points, to account for unevenly sampled data, different exposure times and generally -- the lack of a temporal match between the corresponding points from the two light curves (data sets) that are to be cross-correlated. Thus, the ICCF is being built as a function of the time shift between the curves. The time shift -- $\tau$ (positive or negative), at which ICCF($\tau$) reaches maximum gives the time delay between the curves. Normally, a positive $\tau$ indicates that the first LC is leading the second one. ICCF method is suitable for our case, as the data points are almost equidistant, i.e. there are typically no large missing portions of the intra-night LC to be replaced by straight lines.
\\
\\
The time-delay  results are shown in the lowest panels of Fig. \ref{f1}, \ref{f2}, and \ref{fig:f3}. For the different telescopes, depending on their equipment in use, we compared the optical bands with the highest signal-to-noise and respectively  the LC's providing the highest time resolution. Based on our results we cannot claim any securely detected time delays. Although some indications for such delays are evident (e.g. 21.09.2020, Fig. \ref{f1a}, and 31.08.2021, Fig. \ref{f2b}), first, the results do not appear to be systematic and secondly, the delays are typically within the time resolution, as mentioned above.

\subsection{Polarimetric variability}

The synchrotron radiation is naturally linearly polarized, with a polarization degree that can in theory exceed 50\% under the right circumstances \citep{Ryb1986}. Until recently, the common understanding, however, was that both -- polarization rate and the EVPA are rather stable, or at least -- not highly variable on intra-night time scales. At least during the high-activity episodes, this notion, however, does not seem to hold true. For instance, \citet{Bhat2018} reported significant polarization variability within several hours, during a strong optical micro-flare of S5~0716+714. In a multiwavelength [$\gamma-$ray, optical flux and polarization, plus Very Long Baseline Array (VLBA)] observations of S5 0716+714, \citet{2013ApJ...768...40L} reported a  rapid rotation of the linear polarization coincident with the flare peak in both optical flux and $\gamma-$ray and a new superluminal radio knot appeared essentially at the same time \citep{2015ApJ...809..130C}. In 3C 279, during a multi-wavelength observational campaign from 2008 to 2009, \citet{2010Natur.463..919A} reported a $\gamma-$ray flare coincident with a dramatic change of optical polarization angle.  On another occasion of high $\gamma-$ray activity of 3C 279 in 2011 showed multiple peaks and coincided exactly with a 352$^{\circ}$ rotation of the optical polarization angle and flaring activity at optical bands \citep{2016A&A...590A..10K}. In 3C 454.3, a peculiar optical polarization behaviour was reported during 2009 December 3 -- 12 high multi-wavelength activity. A strong flare peaking in $\gamma-$rays, X-rays and optical/near-infrared almost at the same time showed a  strong anti-correlation between optical flux and degree of polarization, and a large rapid swing in polarization angle of 170$^{\circ}$ \citep{2017MNRAS.472..788G}. There are  several studies of optical polarization variability of blazars on diverse timescales which have shown that virtually all blazars studied are polarimetrically variable on all possible timescales i.e. intra-night, short as well as long \citep[e.g.,][and references therein]{2003A&A...409..857A,2005A&A...442...97A,2011A&A...531A..38A,2011PASJ...63..639I,2016ApJ...831...92B,2017A&A...607A..49S,Mars2021,2022PASJ...74.1041H,Imaz2022}.   
%\citet{Mars2021} showed that virtually all blazars they studied are polarimetrically variable on inter- or intra-night time scales.
\\
\\
The results of our intra-night polarimetric variability are shown in Tab. \ref{tab:polarimetry} and Fig. \ref{fig:f6}. We employed the $\chi^2$ statistics in order  to quantify the reality of the variations in both -- polarization degree and EVPA. For comments on $\chi^{2}$ and comparison among other similar approaches, see e.g. \citet{De2010}. First, we calculated $\chi^{2}$ and the right-tailed probability $P(\chi^{2}, n)$ of the $\chi^{2}$ distribution as:
\\
\\
%\begin{array}{l}
$ \chi^{2}=\sum_{i=1}^{N}\frac{(y_{i}-f(x_{i}))^{2} }{\sigma_{i}^{2}} $\\
$P(\chi^{2}, n)=\frac{2^{-n/2}}{\Gamma(n/2)} \chi^{n-2} e^{-\chi^2/2} $
%\end{array}
\\
\\
We used the weighted-average $\chi^{2}$ approach, as described by \citet[and the references therein]{De2010},  i.e. $f(x)= \langle y \rangle$ is the weighted average of the measured values. Then we used the chi-square value to calculate the probability that the deviations are not due to random variations, i.e. calculated $1-P(\chi^{2}, n)$.  Here $n$ is the degrees of freedom, in our case $n=N-1$, where $N$ is the number of observations. \\
\\
As seen from the results, at least one of these parameters showed statistically significant changes during virtually every night of monitoring. The most impressive cases include the night of 13.09.2021 (Fig. \ref{f6e}), when EVPA changed by about 50 degrees  over 8--9 hours, the night of 11.08.2021, when the polarization rate changed by about 5\% within a few hours, etc. Note that changes  in the polarization parameters do not seem to be related to the changes of the overall flux, i.e. significant optical changes are often not accompanied by polarimetric changes of a similar scale (e.g. the night of 09.08.2021, Fig. \ref{f6b}) and vice versa.

\subsection{The ``\textit{r.m.s.}-flux" relation}

\begin{figure*}
	\includegraphics[width=180mm]{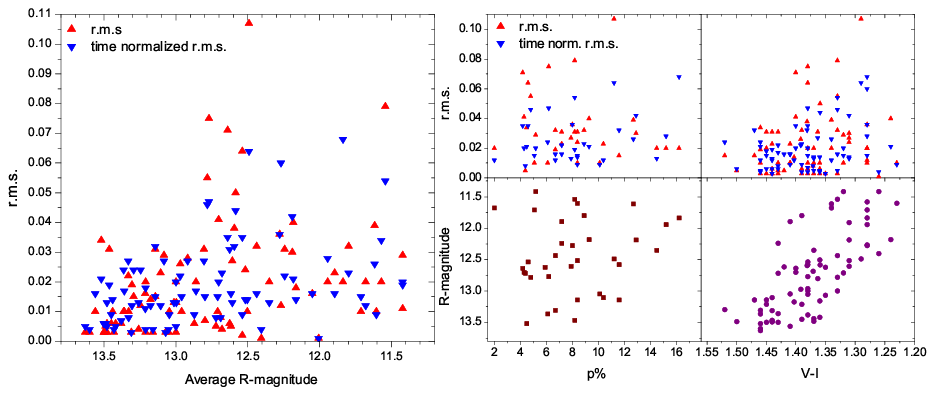}
	\caption{Statistical dependences on intra-night time scale. Left panel: The relation between the variability rate (r.m.s.) and the average \textit{R}-band magnitude. Clearly a tendency for higher fractional variability during episodes of higher flux levels is evident. The right panel shows relations among the flux, the color, the r.m.s. and the polarization degree. No clear dependencies are seen, except for the color-magnitude relation (BWB), which is expected for such types of blazars. }
	\label{f7}
\end{figure*}

The large amount of data acquired during the 2020-2021 and the previous campaigns, allowed studying the "fractional variability -- average flux" or "root-mean-square (\textit{rms})\footnote{Defined as $\sigma_{\rm rms} = \sqrt{\frac{1}{N-1}\sum_{i=1}^{N} (m_{i}-\langle m \rangle)^{2} - \langle \sigma_{\rm phot} \rangle^{2}}$, where $m_{i}$ are the individual magnitude measurements, $\langle m \rangle$ is the average magnitude, $N$ is the number of data points, and $\langle \sigma_{\rm phot} \rangle$ is the average photometric uncertainty. $\sigma_{\rm rms}$ is taken to be zero if the expression under the square root happens to be negative.}
- flux" relation on intra-night time scales. This relation shows how variable an object is at different flux levels. It has been shown for many accreting object, that their variability is flux-dependent \citep[e.g.][and the references therein]{Bach2017}. Fig. \ref{f7} shows the relation between the $rms$ and the average flux we found for BL Lacertae, as well as the relations among other measurable, like color, average polarization degree, etc. Since the typical blazar variability structure function \citep[\textit{SF}; ][]{Sim1985}  is still rising on intra-night time scales, it is justifiable to use in addition a time-normalized $rms$, i.e. $rms(\langle t \rangle/t_{\rm tot})$, where $t_{\rm tot}$ is the time duration of the data set for the corresponding night and $\langle t \rangle$ is the average duration of all data sets, which in our case is 4.05 hours \citep[more details in][]{Bach2015, Bach2016, Bach2017}. Thus, the influence of the length of the observation on the $rms$ value will be minimized.  
Fig. \ref{f7} shows, however, that there is no significant difference in which one of those two parameters are used. Both of them show a slight tendency to increase with the average flux. Similar results have been shown to hold true for other blazars as well; CTA 102 \citep{Bach2017}, S4 0954+65 \citep{Bach2016}, etc. On the other hand, neither the average flux level nor the $rms$ seems to correlate significantly with the average polarization degree.  A peculiar result for BL Lacertae was found in its optical light curve where the flux strongly anti-correlated with the degree of optical polarization while the angle of polarization stayed essentially unchanged \citep{2014ApJ...781L...4G}. To explain this peculiar behaviour of BL Lacertae, \citet{2014ApJ...781L...4G} used the AGN model within the framework of a shock wave propagating along a helical path in the blazar’s jet which can explain the variety of flux and polarization patterns in blazars light curves \citep{2008Natur.452..966M,2013ApJ...768...40L}. 
%\citep[Fig. \ref{f7}; see however][for a different conclusion]{2014ApJ...781L...4G}. 

\begin{subfigures}
\label{fig:f6}
\begin{figure*}
	\includegraphics[width=180mm]{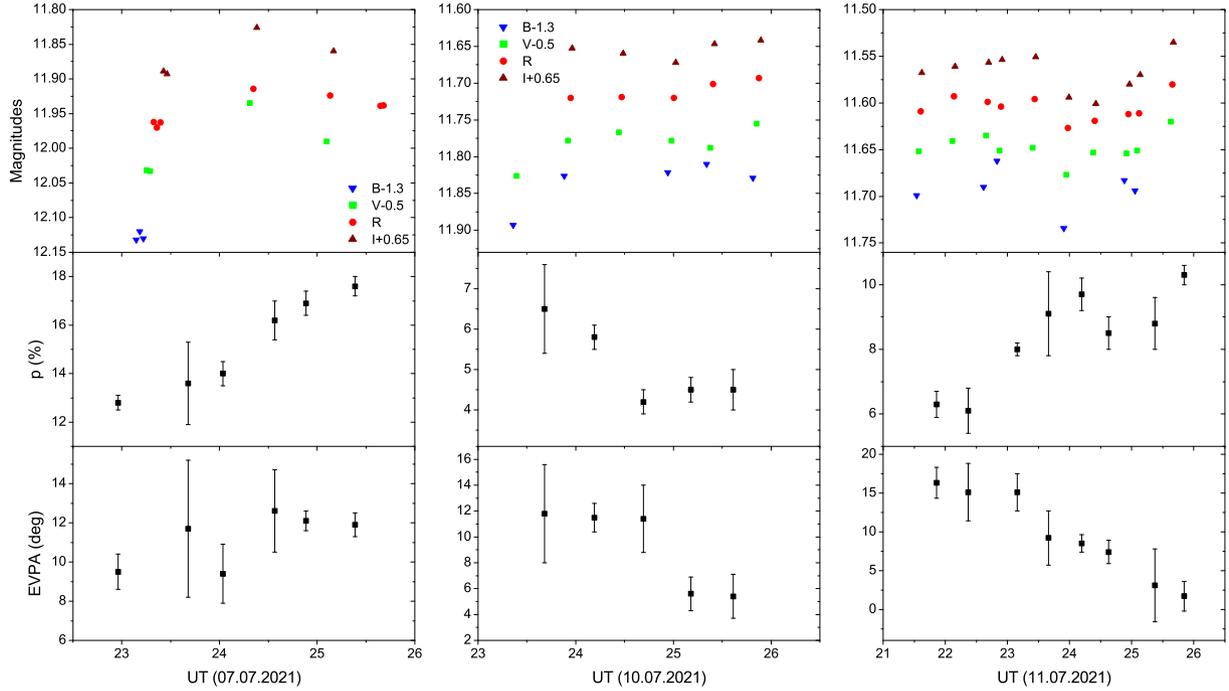}
	\caption{Intra-night multicolor and \textit{R}-band polarimetric observations with the 60cm Belogradchik telescope. \textit{BVRI} light curves (top panel) are shown together with polarization rate (middle panel) and EVPA (bottom panel) for each night. Significant variability in both flux and polarimetric parameters can be seen. These variations do not seem to be correlated in any way.}
	\label{f6a}
\end{figure*}

%\addtocounter{figure}{-1}
\begin{figure*}
	\includegraphics[width=180mm]{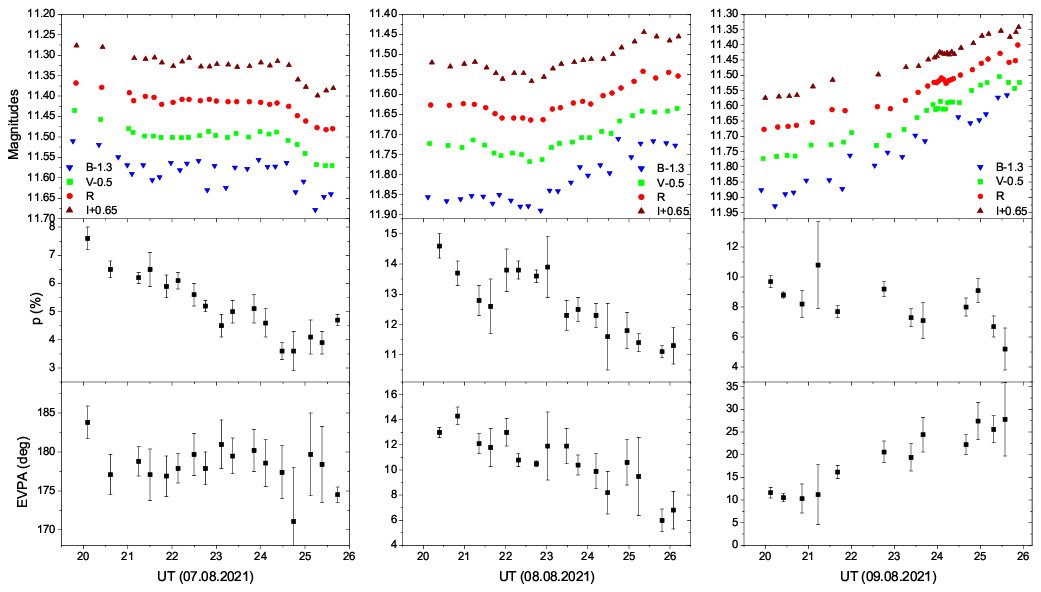}
	\caption{See Figure \ref{f6a}.}
	\label{f6b}
\end{figure*}

%\addtocounter{figure}{-1}
\begin{figure*}
	\includegraphics[width=180mm]{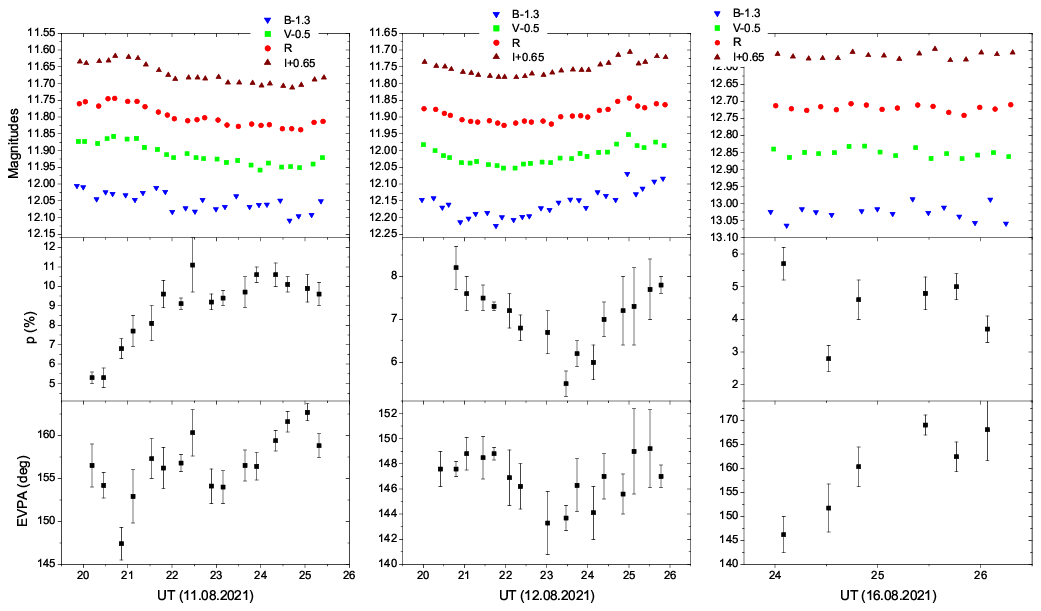}
	\caption{See Figure \ref{f6a}.}
	\label{f6c}
\end{figure*}

%\addtocounter{figure}{-1}
\begin{figure*}
	\includegraphics[width=180mm]{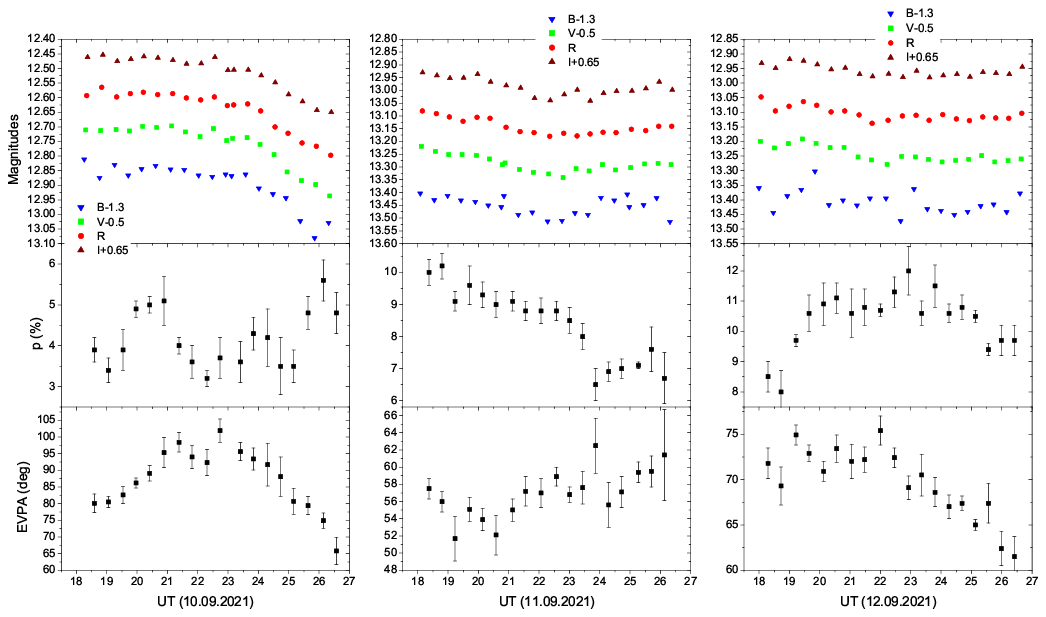}
	\caption{See Figure \ref{f6a}.}
	\label{f6d}
\end{figure*}

%\addtocounter{figure}{-1}
\begin{figure*}
	\includegraphics[width=180mm]{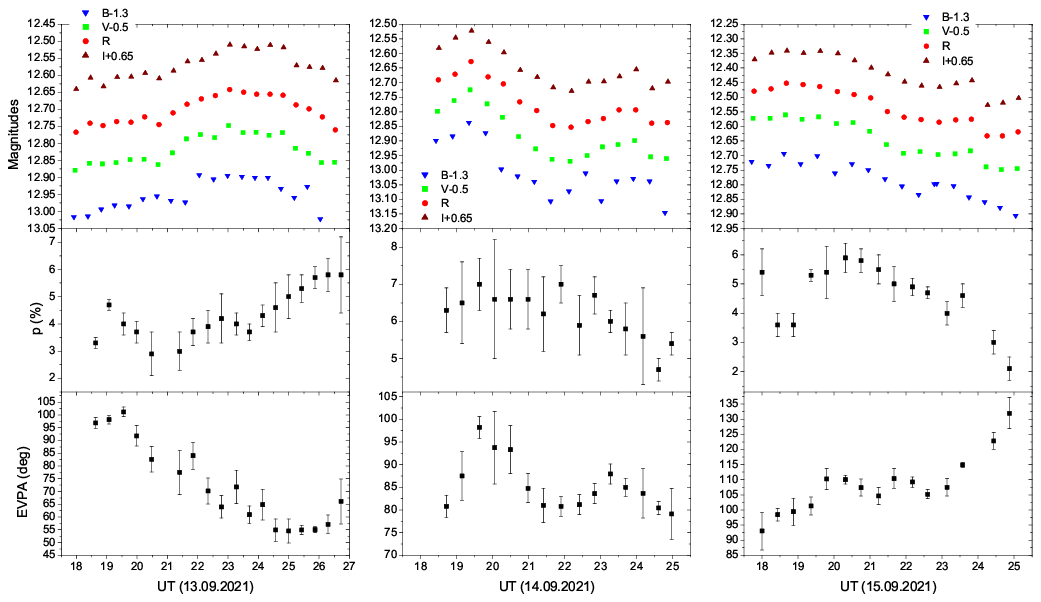}
	\caption{See Figure \ref{f6a}.}
	\label{f6e}
\end{figure*}
\end{subfigures}

\section{Discussion}
Though intra-night variability is now a characteristic feature of identifying blazars, a
detailed explanation of this intra-night variability, often being as rapid as tenths of magnitude in a matter of minutes, is still under debates. Such short characteristic times limit the size of the active region that produces the observed emission and suggest the presence of an ensemble of relatively small emitting regions or blobs. There are %\sout{two} 
major intrinsic reasons %\sout{(microlensing is considered unlikely),} 
able to account for the observed blazar variability\footnote{Here we consider unlikely that the accretion disk is able to account for the observed variations as only jet dominated objects (like blazars), where most of the emission is generated in the jet show such violent variability.}: 

\begin{itemize} 
\item {\bf Geometrical reasons} (i.e. the change of the Doppler factor, $\delta$, of the emitting region). A  geometrical origin of the variability could be a compact region of enhanced emission (or blob) moving helically in the jet \citep{2015ApJ...805...91M,2017MNRAS.465..161S}. 
 The observed-to-emitted flux ratio $F \propto\delta^{3+\alpha}$, where $\delta= 1 / {\Gamma(1-\beta \cos\theta_{obs})}$, the bulk Lorentz factor $\Gamma\simeq10-20$,  $\alpha\simeq1$ is the spectral index, and $\theta_{obs}$ is the angle of the blob with respect to our line of sight \citep[e.g.,][and references therein]{1979ApJ...232...34B,Urry1995,1995ARA&A..33..163W,2019ARA&A..57..467B}.The change of $\delta$ does not change the intrinsic  luminosity of the object, but can significantly modify the emission towards the observer. The $\delta$ is related to the viewing angle ($\theta_{obs}$) of the blob which changes the Doppler boosting and can result in a significant change in the observed flux. The ($\theta_{obs}$) of the blob to the observer's line of sight is given as \citep{2021MNRAS.501...50S,2022MNRAS.510.3641R}
\begin{equation}    
\cos \ \theta_{obs}(t) = \sin\phi \ \sin\psi \ \cos(2\pi t/P_{obs}) + \cos\phi \ \cos\psi 
\end{equation}
here P$_{obs}, \ \phi, \ and \ \psi$ are observed timescale or period of QPO (if detected), pitch angle of the helix, and angle of the axis of the jet with respect to the observer's line of sight, respectively. By substituting the value of $\cos \ \theta_{obs}(t)$ from equation (5) in the expression for $\delta$, we get 
\begin{equation}
F \propto \frac{F^{'}}{\Gamma^{3}(1+S)^{3}} \left(1 - \frac{\beta C}{1+S} \cos(2\pi t/P_{obs} ) \right)
\end{equation}
here F$^{'}$ is emission in the rest-frame, C $\equiv \cos\phi \cos\psi$, and S $\equiv \sin\phi \sin\psi$. \\
 %, small changes of the direction of motion will lead to significant changes in the observed flux. Such direction changes can occur as a result of a curved (helical) magnetic field, along which the emitting blob is moving  \\ 
 %or as a result of turbulence {\bf DO WE MEAN BLOB moving along a curved B or B is curved inside the EMISSION REGION? In the former, we also expect EVPA variation and likely systematically -- as OBSERVED (HELICAL B SCALE much larger scale compared to the spatial extension of the blob).}. \\
\item {\bf Evolution} (acceleration, energy loss due to emission) of the emitting particles. The energy input can be due to a twisted magnetic field, coupled with a rapidly rotating black hole or disk \citep{Blan1977, Blan1982}, 
%(Blandford \& Znajek, 1977; Blandford \& Payne, 1982), 
standing or moving shocks, magnetic reconnections, etc. The losses due to both -- synchrotron and inverse-Compton emission will scale with the particles' energy as $\dot{\gamma}\propto-\gamma^{2}$ \citep{Ryb1986}, implying that the more energetic particles will lose energy faster,  i.e. the characteristic cooling time, $\tau$, will scale as $\tau\propto1/\gamma$. 
\end{itemize}

\noindent
A realistic variability scenario would probably include both of the above-mentioned mechanisms, as well as the presence of more than one actively emitting region at the same (observer’s frame) time. In fact, SED studies \citep{Prin2021,2022MNRAS.513.4645S}, clearly establish the presence of an additional HSP-like (or HBL-like) emission component during this historical activity. Thus, at least, there are two dominant emission components (could be two regions) actively driving the observed behavior. Also, in general, HSP/HBL have relatively low optical polarization degree \citep[e.g.][]{2016ApJ...833...77I,2016MNRAS.463.3365A} compared to LSP and thus, the relatively low polarization degree is likely an outcome of this additional HSP component. We will briefly discuss if and how much our findings, regarding BL~Lacertae, corroborate the existing models. \\
\\
In our observations, we found gradual or rapid rotation of polarization angle EVPA (up to 50 degrees) which can reflect a non-axis symmetric magnetic field \citep{1985ApJ...289..188K}. We rule out the possibility of gradual or rapid change in EVPA due to a uniform axially symmetric, straight, and matter-dominated jet because any  compression of the jet plasma by a perpendicular shock moving along the jet and viewed at a small constant angle to the jet axis would significantly change the degree of polarization $P$, but not result in a gradual or rapid change of EVPA \citep{1985ApJ...289..188K}. According to the another model, the plane of EVPA rotation lies in the projection of plane formed by line of sight and the velocity field on the sky and so the rotation in EVPA can be due to the result from changes in magnetic field topology or Doppler boosting or a combination of both of these \citep{2017MNRAS.467.3876L}. \citet{2014ApJ...780...87M} presented a model called TEMZ (turbulent extreme multi-zone) for variability of flux and polarization of blazars in which turbulent plasma flowing at a relativistic speed down a jet crosses a standing conical shock. According to this model, the randomness in the magnetic field direction in different turbulent cells can cause observed rotations in EVPA, but it probably does not explain the systematic rotations observed in EVPA in the present study of BL Lac. In another model in which the combination of helical magnetic field, straight jet and motion of blob along a straight line can give rise to the rotation of EVPA $\geq$180$^{\circ}$ \citep{2014ApJ...789...66Z,2015ApJ...804...58Z}. 
%Moving blobs along an ordered (e.g. helical) magnetic field most likely will produce EVPA rotation in the same direction \citep{Mars2021}, which is not what we observe in our data. Instead, our results indicate large (up to 50 degrees) and rapid changes of EVPA in any direction but mostly systematic (ref Fig. \ref{fig:f6}). A moving shock will compress the magnetic field perpendicular to the shock direction, which presumably is aligned with the jet, leading to EVPA confined within close proximity to the jet direction \citep[around 10$^{\circ}$,][]{2013MNRAS.436.1530R}. Again, such confinement is not (always) supported by our data. Our results seem to imply the presence of many emitting blobs (active regions) moving in a disordered (turbulent) magnetic field. A similar scheme (TEMZ model) was proposed by \citet{2014ApJ...780...87M} to account for blazars' fast variability and polarization. On the other hand, the Kelvin-Helmholtz turbulence is suppressed if $B>B{\rm c}\simeq10^{-3} n_{\rm e}^{1/2} {\rm G}$ \cite{1995Ap&SS.234...49R}, which is typically fulfilled under normally assumed electron densities, taking into account the BL Lacertae magnetic field of a few G \citep{2022MNRAS.513.4645S}, thus making the development of the turbulence unlikely. \\
%\\
The relatively low polarization degree, compared to the synchrotron theoretical maximum of 75\% \citep{Ryb1986} also suggests the presence of different emitting regions of number $N$ with a randomly oriented magnetic field, as $\langle p \rangle\simeq p_{\mathrm max}/\sqrt{N}$. However, it will be difficult to interpret the outburst with the increase of $N$, as there seems to be no anticorrelation between the flux and the polarization degree (Fig. \ref{f7}). \\
\\
 A more generalized optical flux and polarization variability of BL~Lacertae can be explained as \citet{2008Natur.452..966M} modeled the optical flux, polarization variability with a large swing in EPVA observed in BL~Lacertae in 2005. The results were explained in terms of a shock wave leaving the environs of the central supermassive black hole and propagating down only a portion of the jet’s cross-section. In this case, the disturbance follows a spiral path in a jet that is accelerating and becoming collimated. \citet{2013ApJ...768...40L} have extended \citet{2008Natur.452..966M} model for explaining optical polarization properties in a multi-wavelength outburst of BLL S5~0716+714. They allowed for variations in
the bulk Lorentz factor, $\Gamma$, and keep other parameters e.g. jet viewing angle, the temporal evolution
of the outburst, shocked plasma compression ratio, $k$, spectral
index $\alpha$, and a pitch angle of the spiral motion. In such cases, they found that a wide variety of flux and polarization behaviors still could be reproduced \citep{2013ApJ...768...40L}. The optical flux, polarization and EVPA variation may be incorporated into shock-in-spiral-jet.   \\
\\
Recently, \citet{Jors2022} have reported a $\sim 13$ hours QPO during this historical high state. They showed that the broad variation trends in the observed optical flux and polarization -- degree and EVPA both as well gamma-ray flux can be reproduced by current-driven kink instabilities near a recollimation zone. This instability requires a strong toroidal magnetic field. Our optical flux and polarization observation too show similar trends and thus, this scenario could be a plausible explanation. However, it is apparent that additional contributions are required to explain many of the observed trends. \\
\\
Except on long time scales, when BLLs normally demonstrate bluer-when-brighter behavior \citep[e.g.,][and references therein] {2006A&A...450...39G,Gaur2012,Bhat2018}, BL Lacertae shows small but detectable color changes even on intra-night scales (Fig. \ref{fig:f3}). Here too, on intra-night scales, we observe BWB behaviour. Such changes can either imply rapid evolution of the relativistic particles within an emitting region (cell) or rapid change of the Doppler factors of many cells, allowing different cells to have a major contribution at different times. Since these cells will have perhaps different SED’s, the overall results can rapid change in color in addition to the rapid brightness change. On the other hand, if the evolution (e.g. the energy loss) plays a major role in short-term variability, one may estimate the co-moving magnetic field, $B$, by requiring the cooling time to be shorter than the fastest variability time, $t_{cool} < \frac{\delta}{(1+z)} t_{var}$, 
where $t_{\rm var}\simeq \langle F \rangle /|dF/dt| \simeq 1/|dm/dt|$.
The fastest variations we observed (e.g. 21.09.20, Fig. \ref{f1a}, 28.08.21, Fig. \ref{f2a}) give $t_{var} \simeq 7-10$ h or 30-40 ks. Thus, for the optical region it can be shown \citep{2021ApJS..253...10F}, that the condition above leads to $B > 10 (t_{\mathrm var}, ks)^{-2/3} \mathrm G$, or $B > 0.1 \mathrm G$.
\\
\\
Since the highest energy particles will lose energy faster, if the energy input/loss drives primarily the optical variability, detectable time delays between the bands might be expected. 
Our data, presented here, however, do not support the presence of different from zero delays between the optical bands. This result is not really surprising; many researchers reported either zero (within the errors) optical band lags or nonzero such only in rare cases \citep{2012AJ....143..108W} -- [S5~0716+714]; \citep{2011A&A...531A..90Z,Bach2011} -- [3C~454.3]; \citep{Zhai2012,Bhat2018,Fang2022} -- [BL Lacertae]; \citep{Papa2003,Papa2004,Bach2015} -- [S4~0954+65]; \citep{Bach2017} -- [CTA~102], etc. Note that in some cases (e.g. 18-20.10.20, Fig. \ref{f1c}), due to the overall trends in all bands, the ICCF appears to be very flat-topped and close to 1 during a significant period of time. In such cases, where no sharp peak is present, random, perhaps uncertainty-driven deviations in the light curve can create spurious ICCF peaks of non-zero lags, which we do not consider real-time delays between bands. Not finding time delays, if further confirmed with higher accuracy and cadence observations will be indicative for the changing Doppler factor as the primary driver of the blazar short-term variability. \\
\\
Last, but not the least we address the tendency that blazar's intra-night variability appears to be stronger during a high brightness state. Similar results we have already reported for S4 0954+65 and CTA~102 \citep{Bach2015, Bach2016, Bach2017}. It appears indeed to be just a tendency; however increasing the number of actively emitting regions can increase the brightness, but not the variability. As the brightness will increase $\propto N$, the variations will drop as $1/\sqrt{N}$ at the same time, which is not what we observe for any of these objects, including the BL~Lacertae. The highly non-linear response of the observed flux to the (slowly changing) Doppler factor of the emitting blob can account for the observed positive "rms-flux" correlation.

\section{Conclusions}
The blazar BL~Lacertae has gone through an unprecedented high state of brightness in recent years. We used several telescopes from different observatories throughout the world to monitor its intra-night variability during this event, mostly during the years 2020 --2021. Our goal was to study on the intra-night time scales the optical flux variations, color changes, inter-band time delays, and polarimetric variability (polarization degree and EVPA changes). We found significant variations in all the parameters we studied, including fast changes in the polarization rate and dramatic changes in the EVPA. To the best of our knowledge, this is the first time a blazar has been studied in such details, especially what concerns the optical polarimetry on the intra-night time scales. Our best assumption is that the changing Doppler factor of an ensemble of emitting regions that happen to be on a close alignment with the line of sight, while travelling along the curved jet can be the primary candidate to reproduce the observed picture. Further intra-night variability studies of blazars, especially including polarimetry are highly encouraged and might be essential to resolve the problem.

\section{Acknowledgement}
This research was partially supported by the Bulgarian National Science Fund of the Ministry of Education and Science under grants KP-06-H28/3 (2018), KP-06-H38/4 (2019), KP-06-KITAJ/2 (2020) and KP-06-PN-68/1 (2022). Financial support from the Bulgarian Academy of Sciences (Bilateral grant agreement between BAS and SANU) is gratefully acknowledged. TT acknowledges financial support from the Department of Science and Technology (DST), Government of India (GoI), through INSPIRE fellowship grant No. DST/INSPIRE Fellowship/2019/IF190034. ACG is partially supported by Chinese Academy of Sciences (CAS) President’s International Fellowship Initiative (PIFI) (grant no. 2016VMB073). PK acknowledges financial support from DST, GoI, through the DST-INSPIRE faculty grant (DST/INSPIRE/04/2020/002586). HG acknowledges the financial support from the DST, GoI, through DST-INSPIRE faculty award IFA17-PH197 at ARIES, Nainital, India. GD, OV, and MS acknowledge the observing
and financial grant support from the Institute of Astronomy and Rozhen
NAO BAS through the bilateral SANU-BAN joint research project GAIA CELESTIAL
REFERENCE FRAME (CRF) AND THE FAST VARIABLE ASTRONOMICAL OBJECTS
(2020-2022, leader is G.Damljanovic), and support by the Ministry of
Education, Science and Technological Development of the Republic of
Serbia (contract No 451-03-68/2022-14/200002). The Skinakas Observatory is a collaborative project of the University of Crete, the Foundation for Research and Technology -- Hellas, and the Max-Planck-Institut f\"ur Extraterrestrische Physik. J.H. Fan acknowledges the support from the NSFC (NSFC U2031201, NSFC 11733001, U2031112), Scientific and Technological Cooperation Projects (2020-2023) between the People's Republic of China and the Republic of Bulgaria. Thanks are due to the anonymous referee for the careful reading of the manuscript and the meaningful suggestions that helped to improve this paper.

\section*{Data Availability}
Data presented in the paper may be provided one year after publication. For the data, the request should be made to the lead author of the paper.

\appendix
\section{OBSERVATION LOG}

% \restartappendixnumbering
\setcounter{table}{0}
\begin{table*}
	\caption{Observational log} 
	\label{appendix:A1} 
	\centering 
	%\resizebox{\textwidth} {!}{ 
	\begin{tabular}{lrcccccc} 	\hline\hline 
		JD	&	Evening	&	Telescope	&	Duration	&	Filters	&	$<R>$	&	$\sigma( R )$	&	$<V-I>$	\\\hline 															
		2458011.35	&	14.9.17	&	B60	&	4.6	&	BVRI	&	13.150	&	0.020	&	1.45	 \\
		2458012.35	&	15.9.17	&	B60	&	4.0	&	BVRI	&	13.250	&	$<$0.005	&	1.44	 \\
		2458341.46	&	10.8.18	&	B60	&	2.7	&	BVRI	&	12.900	&	0.010	&	1.38	 \\
		2458343.44	&	12.8.18	&	B60	&	2.5	&	BVRI	&	13.500	&	0.010	&	1.41	 \\
		2458344.43	&	13.8.18	&	B60	&	4.0	&	BVRI	&	12.940	&	0.020	&	1.38	 \\
		2458363.39	&	1.9.18	&	B60	&	2.3	&	BVRI	&	12.940	&	$<$0.005	&	1.33	 \\
		2458364.46	&	2.9.18	&	B60	&	4.7	&	BVRI	&	13.010	&	$<$0.005	&	1.35	 \\
		2458367.48	&	5.9.18	&	B60	&	3.3	&	BVRI	&	12.990	&	$<$0.005	&	1.38	 \\
		2458402.27	&	10.10.18	&	B60	&	2.0	&	BVRI	&	13.440	&	$<$0.005	&	1.45	 \\
		2458403.38	&	11.10.18	&	B60	&	3.4	&	BVRI	&	13.540	&	$<$0.005	&	1.46	 \\
		2458404.27	&	12.10.18	&	B60	&	2.5	&	BVRI	&	13.570	&	$<$0.005	&	1.46	 \\
		2458405.36	&	13.10.18	&	B60	&	2.2	&	BVRI	&	13.430	&	$<$0.005	&	1.44	 \\
		2458406.28	&	14.10.18	&	B60	&	2.4	&	BVRI	&	13.300	&	0.010	&	1.43	 \\
		2458407.30	&	15.10.18	&	B60	&	1.9	&	BVRI	&	13.440	&	$<$0.005	&	1.43	 \\
		2458408.27	&	16.10.18	&	B60	&	1.9	&	BVRI	&	13.420	&	$<$0.010	&	1.44	 \\
		2458428.27	&	5.11.18	&	R200	&	3.4	&	BR	&	13.270	&	0.022	&	--	 \\
		2458676.36	&	11.7.19	&	B60	&	2.8	&	BVRI	&	13.100	&	$<$0.005	&	1.36	 \\
		2458721.46	&	25.8.19	&	B60	&	2.7	&	BVRI	&	13.420	&	$<$0.005	&	1.37	 \\
		2458722.40	&	26.8.19	&	B60	&	4.7	&	BVRI	&	13.360	&	$<$0.010	&	1.36	 \\
		2458723.34	&	27.8.19	&	B60	&	2.8	&	BVRI	&	13.380	&	0.010	&	1.37	 \\
		2458724.48	&	28.8.19	&	B60	&	2.8	&	BVRI	&	13.390	&	$<$0.010	&	1.39	 \\
		2458725.46	&	29.8.19	&	B60	&	3.3	&	VRI	&	13.320	&	$<$0.010	&	1.38	 \\
		2458754.33	&	27.9.19	&	B60	&	3.0	&	VRI	&	13.150	&	$<$0.005	&	1.39	 \\
		2458783.25	&	26.10.19	&	B60	&	2.8	&	Unfiltered	&	13.370	&	$<$0.005	&	--	 \\
		2459079.47	&	17.8.20	&	R200	&	4.1	&	BR	&	12.050	&	0.016	&	--	 \\
		2459104.50	&	11.9.20	&	V140	&	3.0	&	BVRI	&	12.503	&	0.010	&	1.36	 \\
		2459112.27	&	19.9.20	&	B60	&	7.7	&	BVRI + pol R	&	12.182	&	0.040	&	1.24	 \\
		2459112.39	&	19.9.20	&	R200	&	5.0	&	BR	&	12.164	&	0.018	&	--	 \\
		2459113.29	&	20.9.20	&	B60	&	8.4	&	BVRI + pol R	&	12.606	&	0.027	&	1.31	 \\
		2459114.28	&	21.9.20	&	B60	&	6.7	&	BVRI + pol R	&	12.490	&	0.107	&	1.29	 \\
		2459115.30	&	22.9.20	&	B60	&	7.2	&	BVRI + pol R	&	12.517	&	0.024	&	1.31	 \\
		2459116.35	&	23.9.20	&	B60	&	4.1	&	BVRI + pol R	&	12.278	&	0.036	&	1.28	 \\
		2459117.28	&	24.9.20	&	B60	&	8.4	&	BVRI + pol R	&	12.438	&	0.032	&	1.33	 \\
		2459132.36	&	9.10.20	&	V140	&	0.8	&	BVRI	&	12.269	&	0.012	&	1.28	 \\
        2459133.40	&	10.10.20	&	V140	&	0.9	&	BVRI	&	12.404	&	0.001	&	1.26	 \\
		2459134.25	&	11.10.20	&	B60	&	4.8	&	BVRI + pol R	&	12.780	&	0.055	&	1.33	 \\
		2459137.36	&	14.10.20	&	B60	&	2.6	&	BVRI + pol R	&	12.710	&	0.010	&	1.32	 \\
		2459140.17	&	17.10.20	&	A104	&	5.8	&	R	&	13.212	&	0.016	&	--	 \\
		2459141.29	&	18.10.20	&	B60	&	6.5	&	BVRI + pol R	&	13.464	&	0.031	&	1.44	 \\
		2459142.29	&	19.10.20	&	B60	&	6.5	&	BVRI + pol R	&	13.519	&	0.034	&	1.46	 \\
		2459142.31	&	19.10.20	&	V60	&	3.3	&	BVRI	&	13.488	&	$<$0.005	&	1.50	 \\
		2459143.29	&	20.10.20	&	B60	&	6.7	&	BVRI + pol R	&	13.306	&	0.019	&	1.46	 \\
		2459143.27	&	20.10.20	&	V60	&	2.5	&	BVRI	&	13.295	&	0.015	&	1.52	 \\
		2459144.24	&	21.10.20	&	B60	&	1.7	&	BVRI + pol R	&	13.363	&	0.010	&	1.46	 \\
		2459144.36	&	21.10.20	&	R200	&	4.8	&	BR	&	13.329	&	0.010	&	--	 \\
		2459145.26	&	22.10.20	&	B60	&	4.6	&	BVRI + pol R	&	13.046	&	0.010	&	1.41	 \\
		2459145.35	&	22.10.20	&	R200	&	4.1	&	BR	&	13.013	&	0.013	&	--	 \\
		2459149.13	&	26.10.20	&	A104	&	4.3	&	R	&	12.917	&	0.028	&	--	 \\
		2459150.11	&	27.10.20	&	A104	&	4.0	&	R	&	12.997	&	0.013	&	--	 \\
		2459155.05	&	1.11.20	&	A104	&	1.1	&	R	&	13.260	&	0.004	&	--	 \\
		2459160.13	&	6.11.20	&	A104	&	4.4	&	R	&	13.082	&	0.029	&	--	 \\
		2459161.44	&	7.11.20	&	B60	&	1.9	&	BVRI + pol R	&	13.142	&	0.015	&	1.47	 \\
		2459172.05	&	18.11.20	&	A104	&	0.8	&	R	&	12.679	&	0.004	&	--	 \\
		2459181.07	&	27.11.20	&	A104	&	2.0	&	R	&	12.797	&	0.007	&	--	 \\
		2459188.08	&	4.12.20	&	A104	&	2.6	&	R	&	12.872	&	0.006	&	--	 \\
		2459189.08	&	5.12.20	&	A104	&	2.1	&	R	&	13.246	&	0.012	&	--	 \\
		2459195.07	&	11.12.20	&	A104	&	0.8	&	R	&	12.626	&	0.006	&	--	 \\
		2459203.12	&	19.12.20	&	A104	&	1.0	&	R	&	12.536	&	0.002	&	--	 \\
		2459204.30	&	20.12.20	&	V60	&	1.3	&	BVRI	&	12.615	&	$<$0.005	&	1.42	 \\\hline 
		\end{tabular}
\end{table*}

\clearpage
\setcounter{table}{0}
\begin{table*}
\caption{Continued} 
\centering 	
	%\resizebox{\textwidth} {!}{ 
	\begin{tabular}{lrcccccc} \hline\hline 
		JD	&	Evening	&	Telescope	&	Duration	&	Filters	&	$<R>$	&	$\sigma( R )$	&	$<V-I>$	\\\hline
		2459403.51	&	7.7.21	&	B60	&	2.9	&	BVRI + pol R	&	11.944	&	0.020	&	1.28	 \\
		2459404.51	&	8.7.21	&	B60	&	2.9	&	BVRI + pol R	&	12.188	&	0.030	&	1.31	 \\
		2459406.52	&	10.7.21	&	B60	&	2.6	&	BVRI + pol R	&	11.711	&	0.010	&	1.28	 \\
		2459407.49	&	11.7.21	&	B60	&	4.3	&	BVRI + pol R	&	11.605	&	0.010	&	1.23	 \\
		2459408.42	&	12.7.21	&	B60	&	1.2	&	BVRI + pol R	&	11.835	&	0.020	&	1.28	 \\
		2459429.35	&	2.8.21	&	R200	&	2.4	&	BR	&	11.570	&	0.020	&	--	 \\
		2459429.54	&	3.8.21	&	R200	&	2.4	&	BR	&	11.420	&	0.011	&	--	 \\
		2459434.45	&	7.8.21	&	B60	&	5.8	&	BVRI + pol R	&	11.420	&	0.029	&	1.32	 \\
		2459435.46	&	8.8.21	&	B60	&	6.0	&	BVRI + pol R	&	11.615	&	0.039	&	1.34	 \\
		2459436.50	&	9.8.21	&	B60	&	5.9	&	BVRI + pol R	&	11.542	&	0.079	&	1.33	 \\
		2459437.45	&	10.8.21	&	B60	&	6.5	&	BVRI + pol R	&	11.678	&	0.020	&	1.34	 \\
		2459438.43	&	11.8.21	&	B60	&	5.5	&	BVRI + pol R	&	11.797	&	0.032	&	1.39	 \\
		2459439.44	&	12.8.21	&	B60	&	6.0	&	BVRI + pol R	&	11.893	&	0.023	&	1.41	 \\
		2459440.42	&	13.8.21	&	B60	&	5.8	&	BVRI + pol R	&	12.242	&	0.031	&	1.43	 \\
		2459443.54	&	16.8.21	&	B60	&	2.4	&	BVRI + pol R	&	12.720	&	0.005	&	1.43	 \\
		2459453.40	&	26.8.21	&	S130	&	4.3	&	BVRI	&	12.687	&	0.008	&	1.37	 \\
		2459454.40	&	27.8.21	&	S130	&	4.6	&	BVRI	&	12.584	&	0.050	&	1.36	 \\
		2459455.39	&	28.8.21	&	S130	&	4.8	&	BVRI	&	12.593	&	0.038	&	1.38	 \\
		2459456.40	&	29.8.21	&	S130	&	4.8	&	BVRI	&	12.805	&	0.031	&	1.36	 \\
		2459457.39	&	30.8.21	&	S130	&	4.8	&	BVRI	&	12.970	&	0.026	&	1.39	 \\
		2459458.38	&	31.8.21	&	S130	&	4.8	&	BVRI	&	13.176	&	0.014	&	1.40	 \\
		2459468.42	&	10.9.21	&	B60	&	8.2	&	BVRI + pol R	&	12.640	&	0.071	&	1.40	 \\
		2459469.43	&	11.9.21	&	B60	&	8.2	&	BVRI + pol R	&	13.142	&	0.031	&	1.45	 \\
		2459470.43	&	12.9.21	&	B60	&	8.2	&	BVRI + pol R	&	13.105	&	0.023	&	1.44	 \\
		2459471.42	&	13.9.21	&	B60	&	8.2	&	BVRI + pol R	&	12.703	&	0.041	&	1.40	 \\
		2459472.40	&	14.9.21	&	B60	&	6.5	&	BVRI + pol R	&	12.770	&	0.075	&	1.38	 \\
		2459473.40	&	15.9.21	&	B60	&	7.4	&	BVRI + pol R	&	12.536	&	0.064	&	1.38	 \\
		2459490.49  &   2.10.21 &  V140 &  3.9 &  BVRI              &   12.137  &  0.010   &   1.37 \\
        2459520.34	&	1.11.21	&	B60	&	6.5	&	BVRI + pol R	&	12.350	&	0.020	&	1.37	 \\
		2459549.35	&	30.11.21	&	R200	&	3.4	&	BR	&	12.005	&	0.001	&	--	 \\
		2459550.30	&	1.12.21	&	R200	&	4.8	&	BR	&	12.860	&	0.020	&	--	 \\\hline 
		\end{tabular}

{\bf Telescope codes:}\\
B60: 60-cm Cassegrain telescope, Belogradchik Astronomical Observatory, Bulgaria\\
R200: 200-cm RCC telescope, Rozhen National Astronomical Observatory, Bulgaria \\
A104: 104-cm ARIES Sampuranand telescope, Nainital, India\\
S130: 130-cm Skinakas telescope, Crete, Greece \\
V60: 60 cm-telescope, Vidojevica Astronomical Station, Serbia \\
V140: 140-cm telescope, Vidojevica Astronomical Station, Serbia
\end{table*}

\clearpage
\section{IDV RESULTS}

% \restartappendixnumbering
\setcounter{table}{0}
\begin{table*}
	\caption{Results of IDV of BL Lac} 
	\label{appendix:B1} 
	\centering
\begin{tabular}{ccccccccc} 
	   \hline\hline
Obs. date & Obs. start time    & Band &  \multicolumn{2}{c}{ Power enhanced F-test}& \multicolumn{2}{c}{Nested ANOVA} & Variability & $A$\\
          
dd-mm-yyyy &       JD     & & DoF($\nu_1,\nu_2$) & $F_{enh}$ / $F_c$ & DoF($\nu_1,\nu_2$) & F /  $F_c$   & Status & (\%) \\
	    \hline
2017-09-14 & 2458011.26980 & V & 38,76 & 2.27/1.87 &  9,30 & 38.47/3.06 & V & 8.22  \\ 
           & 2458011.26539 & R & 39,78 & 4.12/1.86 &  9,30 & 41.78/3.06 & V & 7.84   \\ 
           & 2458011.26685 & I & 40,80 & 2.26/1.84 &  9,30 & 20.84/3.06 & V & 8.86    \\ 
		
2017-09-15 & 2458012.29265 & V & 29,58 & 1.21/2.05 &  6,21 & 6.62/3.81 & NV  & -- \\ 
           & 2458012.28825 & R & 28,56 & 1.09/2.07 &  6,21 & 1.54/3.81 & NV  & -- \\ 
           & 2458012.28972 & I & 28,56 & 1.38/2.07 &  6,21 & 2.58/3.81 & NV  & --    \\ 
		
2018-08-10 & 2458341.40920 & V & 22,44 & 3.62/2.27 &  5,18 & 17.82/4.24 & V & 8.33\\ 
           & 2458341.40479 & R & 22,44 & 4.65/2.27 &  5,18 & 32.79/4.24 & V & 6.86   \\ 
           & 2458341.40625 & I & 22,44 & 1.19/2.27 &  5,18 & 24.09/4.24 & NV & --   \\ 
		
2018-08-12 & 2458343.39397 & V & 20,40 & 1.98/2.36 &  4,15 & 17.88/4.89 & NV & -- \\ 
           & 2458343.38957 & R & 20,40 & 2.41/2.36 &  4,15 & 3.98/4.89 & NV  & -- \\ 
           & 2458343.39103 & I & 20,40 & 2.24/2.36 &  4,15 & 3.25/4.89 & NV  & -- \\ 
		
2018-08-13 & 2458344.35588 & V & 33,66 & 3.83/1.96 &  7,24 & 21.41/3.49 & V & 13.45 \\ 
           & 2458344.35148 & R & 34,68 & 2.25/1.94 &  8,27 & 24.87/3.25 & V & 7.76 \\ 
	   & 2458344.35294 & I & 33,66  & 2.53/1.96 &  7,24 & 32.39/3.49 & V & 5.44 \\ 
		
2018-09-01 & 2458363.34850 & V & 18,36 & 1.11/2.47 &  4,15 & 4.36/4.89 & NV & -- \\ 
           & 2458363.34404 & R & 18,36 & 0.69/2.47 &  4,15 & 7.30/4.89 & NV & --\\ 
           & 2458363.34552 & I & 18,36 & 0.67/2.47 &  4,15 & 6.21/4.89 & NV & --\\ 
		
2018-09-02 & 2458364.37688 & V & 38,76 & 2.25/1.87 &  9,30 & 14.41/3.06 & V & 15.22 \\ 
           & 2458364.37248 & R & 38,76 & 1.83/1.87 &  9,30 & 34.37/3.06 & NV & -- \\ 
           & 2458364.37394 & I & 37,74 & 1.64/1.89 &  8,27 & 40.13/3.25 & NV & -- \\ 
		
2018-09-05 & 2458367.42068 & V & 27,54 & 0.84/2.10 &  6,21 & 0.93/3.81 & NV & -- \\ 
           & 2458367.41627 & R & 27,54 & 1.04/2.10 &  6,21 & 7.15/3.81 & NV & -- \\ 
	   & 2458367.41774 & I & 27,54 & 0.81/2.10 &  6,21 & 2.24/3.81  & NV & -- \\ 
		
2018-10-10 & 2458402.23031 & V & 17,34 & 0.86/2.54 &  3,12 & 1.52/5.95 & NV & -- \\ 
           & 2458402.22589 & R & 18,36 & 0.66/2.47 &  4,15 & 2.98/4.89 & NV & -- \\ 
           & 2458402.23323 & I & 17,34 & 0.68/2.54 &  3,12 & 1.39/5.95 & NV & -- \\ 
		
2018-10-11 & 2458403.31939 & V & 28,56 & 0.89/2.07 &  6,21 & 3.36/3.81 & NV & -- \\ 
           & 2458403.31499 & R & 28,56 & 1.97/2.07 &  6,21 & 10.57/3.81 & NV & -- \\ 
           & 2458403.31645 & I & 26,52 & 1.08/2.13 &  6,21 & 5.53/3.81 & NV  & -- \\ 
		
2018-10-12 & 2458404.22027 & V & 21,42 & 0.92/2.32 &  4,15 & 5.06/4.89 & NV & -- \\ 
           & 2458404.22172 & R & 20,40 & 0.68/2.36 &  4,15 & 8.48/4.89 & NV & -- \\ 
           & 2458404.22763 & I & 20,40 & 0.57/2.36 &  4,15 & 2.02/4.89 & NV & -- \\ 
		
2018-10-13 & 2458405.32499 & V & 19,38 & 0.75/2.42 &  4,15 & 0.70/4.89 & NV & --\\ 
           & 2458405.32058 & R & 19,38 & 0.82/2.42 &  4,15 & 2.40/4.89 & NV & --\\ 
           & 2458405.32205 & I & 19,38 & 0.68/2.42 &  4,15 & 3.06/4.89 & NV & --\\ 
		
2018-10-14 & 2458406.23311 & V & 18,36 & 1.95/2.47 &  4,15 & 10.9/4.89 & NV & --\\ 
           & 2458406.22870 & R & 19,38 & 2.23/2.42 &  4,15 & 12.61/4.89 & NV & --    \\ 
           & 2458406.23017 & I & 17,34 & 2.90/2.54 &  3,12 & 8.12/5.95 & V & 13.88    \\ 
		
2018-10-15 & 2458407.26002 & V & 15,30 & 1.81/2.70 &  3,12 & 7.80/5.95 & NV & --\\ 
           & 2458407.25561 & R & 15,30 & 2.26/2.70 &  3,12 & 2.72/5.95 & NV & -- \\ 
           & 2458407.25708 & I & 14,28 & 0.75/2.79 &  3,12 & 0.26/5.95 & NV & -- \\ 
		
2018-10-16 & 2458408.22941 & V & 15,30 & 1.99/2.70 &  3,12 & 1.34/5.95 & NV & --\\ 
           & 2458408.23087 & R & 15,30 & 1.27/2.70 &  3,12 & 6.36/5.95 & NV & --\\ 
           & 2458408.23233 & I & 14,28 & 0.64/2.79 &  3,12 & 1.33/5.95 & NV & -- \\ 
		
2019-07-11 & 2458676.44236 & V & 13,26 & 1.23/2.90 &  2,9  & 0.18/8.02 & NV & --\\ 
           & 2458676.43796 & R & 14,28 & 0.87/2.79 &  3,12 & 0.62/5.95 & NV & -- \\ 
           & 2458676.43942 & I & 14,28 & 1.08/2.79 &  3,12 & 3.89/5.95 & NV & --\\ 
		
2019-08-25 & 2458721.43409 & V & 14,28 & 1.37/2.79 &  3,12 & 3.17/5.95 & NV & --\\ 
           & 2458721.42968 & R & 14,28 & 1.01/2.79 &  3,12 & 4.79/5.95 & NV & --    \\ 
           & 2458721.43701 & I & 13,26 & 0.97/2.90 &  2,9  & 3.29/8.02 & NV & --    \\ 
		
2019-08-26 & 2458722.30764 & V & 39,78 & 0.72/1.86 &  9,30 & 4.60/3.06 & NV & --\\ 
           & 2458722.30323 & R & 37,74 & 0.79/1.89 &  8,27 & 14.81/3.25 & NV & --   \\ 
           & 2458722.30469 & I & 37,74 & 0.86/1.89 &  8,27 & 8.35/3.25 & NV & --  \\ 
		
2019-08-27 & 2458723.29235 & V & 24,48 & 1.98/2.20 &  5,18 & 5.58/4.24 & NV & -- \\ 
           & 2458723.28795 & R & 24,48 & 3.70/2.20 &  5,18 & 15.18/4.24 & V & 12.98    \\ 
           & 2458723.28941 & I & 24,48 & 2.62/2.20 &  5,18 & 20.16/4.24 & V & 13.08   \\\hline 
 
           \\
  \end{tabular}
  %\label{tab:resIDV}
\end{table*} 

\clearpage
\setcounter{table}{0}
\begin{table*}
\caption{Continued}
\begin{tabular}{ccccccccc}
	   \hline \hline
Obs. date & Obs. start time & Band & \multicolumn{2}{c}{Power enhanced F-test} &\multicolumn{2}{c}{ Nested ANOVA} & Variability & $A$\\

dd-mm-yyyy & JD & & DoF($\nu_1,\nu_2$) & $F_{enh}$ / $F_c$ & DoF($\nu_1,\nu_2$) & F /  $F_c$ & Status & (\%) \\\hline         
	2019-08-28 & 2458724.42546 & V & 24,48 & 0.28/2.20 &  5,18 & 1.71/4.24 & NV & -- \\ 
           & 2458724.42105 & R & 23,46 & 0.83/2.23 &  5,18 & 2.75/4.24 & NV & --   \\ 
           & 2458724.42251 & I & 24,48 & 0.18/2.20 &  5,18 & 0.58/4.24 & NV & --   \\	
2019-08-29 & 2458725.39609 & V & 27,54 & 1.47/2.10 & 6,21 & 24.8/3.81 & NV & -- \\ 
           & 2458725.39168 & R & 28,56 & 1.66/2.07 & 6,21 & 28.86/3.81 & NV & --   \\ 
           & 2458725.39315 & I & 28,56 & 1.92/2.07 & 6,21 & 39.05/3.81 & NV & --   \\ 
		
2019-09-27 & 2458754.27882 & V & 25,50 & 0.66/2.16 & 5,18 & 1.34/4.24 & NV & --\\ 
           & 2458754.27439 & R & 19,38 & 1.51/2.42 & 4,15 & 4.48/4.89 & NV & --  \\
           & 2458754.27587 & I & 25,50 & 0.37/2.16 & 5,18 & 1.33/4.24 & NV & --   \\
2020-10-26 & 2459149.03856 & R & 489,978  & 15.54/1.19 & 121,366 & 159.59/1.39 & V & 8.99   \\ 
2020-10-27 & 2459150.02883 & R & 577,1154 & 4.23/1.18  & 143,432 & 83.42/1.35  & V    & 6.59  \\ 
2020-11-01 & 2459155.02320 & R & 137,274  & 0.61/1.40  & 33,102  & 6.12/1.85   & NV & --     \\ 
2020-11-06 & 2459160.03965 & R & 553,1106 & 30.48/1.18 & 137,414 & 86.7/1.36   & V & 11.90     \\ 
2020-11-18 & 2459172.02950 & R & 133,266  & 2.68/1.40  & 32,99   & 5.27/1.87   & V & 2.46    \\ 
2020-11-27 & 2459181.02645 & R & 287,574  & 2.49/1.26  & 71,216  & 6.63/1.53   & V & 3.76   \\ 
2020-12-04 & 2459188.03180 & R & 373,746  & 3.07/1.22  & 92,279  & 6.27/1.46   & V & 3.78     \\ 
2020-12-05 & 2459189.04601 & R & 203,406  & 5.66/1.31  & 50,153  & 7.43/1.66   & V & 4.58     \\ 
2020-12-11 & 2459195.05651 & R & 197,394  & 3.82/1.32  & 48,147  & 9.46/1.67   & V & 3.45    \\ 
2020-12-19 & 2459203.09193 & R & 113,226  & 1.08/1.44  & 27,84   & 2.01/1.97   & NV & --    \\		\hline
\\
  \end{tabular}
%  \label{tab:resIDV}
\end{table*}

\end{document}